\documentclass[journal]{IEEEtran}
\usepackage[utf8]{inputenc}
\usepackage{graphicx}
\usepackage{tabularx,booktabs}
\usepackage[T1]{fontenc}
\usepackage[export]{adjustbox}
\usepackage{amsmath,cite}
\usepackage{amssymb}
\usepackage{gensymb}
\usepackage{bbm}
\usepackage{subfig}
\usepackage{listings}
\usepackage{mathtools}

\newtheorem{theorem}{\textbf{Theorem}}

\usepackage{pgfplots}
  \pgfplotsset{compat=newest}
\usepackage{wasysym} 
\usepackage{graphicx}

\usetikzlibrary {positioning}

\usepackage{tikz}
\usetikzlibrary{patterns}
\usepackage{algorithm}
\usepackage{algpseudocode}
\usepackage{nicefrac}
\usepackage{kbordermatrix}

\renewcommand{\kbldelim}{(}
\renewcommand{\kbrdelim}{)}

\usepackage{amsmath}

\begin{document}
%
\title{Spatiotemporal Analysis of Parallelized Computing at the Extreme Edge}
\author{Yasser Nabil, Graduate Student Member, IEEE, Mahmoud Abdelhadi, Sameh Sorour,\\ Hesham ElSawy, Senior Member, IEEE, Sara A. Elsayed, Member, IEEE, and Hossam S. Hassanein, Fellow, IEEE
\thanks{ This research is supported by a grant from the Natural Sciences
and Engineering Research Council of Canada (NSERC)
under grant number: ALLRP 549919-20, and a grant from
Distributive, Ltd.\\ Y.\ Nabil is with the Electrical and Computer Engineering Department, Queen’s University, Kingston, Ontario, Canada. E-mail: \texttt{yasser.nabil@queensu.ca}. \\ 
M. Abdelhadi,  S. Sorour, H.\ ElSawy, and H. S.  Hassanein are with the School of Computing, Queen's University, Kingston, Ontario, Canada. E-mails: \texttt{\{m.abdelhadi, hesham.elsawy\}@queensu.ca, and hossam@cs.queensu.ca}.\\
S. A. Elsayed is with the Department of Computer Science, University of Calgary, Calgary, Alberta, Canada.   E-mail: \texttt{sara.elsayed@ucalgary.ca}\\
  Y.\ Nabil and M. Abdelhadi have contributed equally to this work.}\thanks{This work was presented in part in \cite{abdelhadi2022parallel}.}}

\maketitle

\begin{abstract}
Low-latency computational-task execution can be achieved by leveraging device-to-device offloading and parallel processing over nearby extreme edge devices (EEDs), a paradigm known as extreme edge computing (EEC). However, EEC performance is challenged by device spatial randomness with intermittent wireless connectivity, limited device computing power, time-varying availability, and device failures. This paper introduces a novel spatiotemporal analytical framework for EEC by integrating stochastic geometry with an absorbing continuous-time Markov chain (ACTMC) to capture the interplay between communication and computation. Modeling a large-scale millimeter-wave network, we derive tractable expressions for the average task response delay and the task completion probability under both random and location-aware EED selection. Numerical results quantify the impact of location-awareness and unveil the existence of an optimal task segmentation that minimizes delay, which depends on network parameters and EED capabilities. We also demonstrate that device failures and EED scarcity exacerbate delay, which can be mitigated through a collaborative load-balancing approach between EEC and Multi-Access Edge Computing (MEC) schemes. Simulations and sensitivity analyses validate the proposed framework and offer design insights for optimizing system performance.
\end{abstract}

\begin{IEEEkeywords}
Extreme edge computing (EEC), multi-access edge computing (MEC), task offloading, millimeter-wave communication, stochastic geometry, spatiotemporal modeling.
\end{IEEEkeywords}

%
\IEEEpeerreviewmaketitle

\section{Introduction}

Sixth-generation (6G) networks are anticipated to establish an infrastructure capable of supporting highly interconnected intelligent ecosystems \cite{ma2025artificial,wang2023road}. The anticipated 6G architecture features a diverse, intelligent, and perceptive structure facilitated by robust edge servers and distributed computing facilities \cite{nguyen20216g,letaief2021edge}. This enables a wide range of applications, such as digital twins, remote surgeries, smart cities, autonomous vehicles, industrial autonomy, and the Tactile Internet \cite{ma2025artificial,wang2023road}.
Furthermore, 6G is projected to lead to an increase in device-to-device (D2D) connections, extensive utilization of artificial intelligence (AI), and a surge in the Internet of Things (IoT) services \cite{ma2025artificial,wang2023road,nguyen20216g,letaief2021edge}.   This is expected to trigger an unprecedented increase in data traffic and a corresponding need for extensive computations in the network.

Extensive computation is required for many AI workloads that underpin latency-sensitive IoT applications with strict quality-of-service (QoS) requirements  \cite{hazra2023fog,hua2023edge}. One solution to handle such extensive computations is to utilize cloud computing by offloading tasks to remote data centers. However, cloud computing fails to adequately satisfy latency-sensitive applications due to the distant geographical location of data centers and the huge traffic influx imposed at backhaul links \cite{hazra2023fog,hua2023edge}. Multi-access edge computing (MEC) has emerged as a promising paradigm that can bring computing services closer to end devices, effectively reducing latency and meeting the increasing demands of IoT applications  \cite{pham2020survey}. MEC platforms typically rely on edge servers co-located with or near base stations (BSs) to handle offloaded computational tasks, making efficient task offloading decisions crucial for achieving optimal MEC performance. The process of task offloading in MEC environments is significantly influenced by the availability, accessibility, and resilience of resources  \cite{bagchi2019dependability}. However, managing these factors can be costly and may not be applicable in certain scenarios.  Additionally, the increasing number of devices utilizing MEC has given rise to the unresolved challenge of high congestion.

To overcome these limitations, extreme edge computing (EEC) offers an attractive alternative that leverages the abundant yet underutilized computational resources of IoT devices, referred to as extreme edge devices (EEDs), including smartphones, laptops, and connected vehicles \cite{portilla2019extreme}.
While individual IoT devices possess limited processing power, their collective computational capabilities, when used in parallel, represent a significant untapped resource \cite{el2025proactive}.  In EEC, these devices are harnessed to expand the computational resource pool, facilitate parallel processing, and improve task offloading by bringing the computing service much closer to end users, thus significantly reducing the response delay. Utilizing the abundant and underutilized computational resources of EEDs can also disrupt the dominance of traditional cloud service providers and network operators, fostering a more decentralized and democratized edge computing ecosystem with substantial advantages.

Despite its promising advantages, the EEC architecture faces several unique and interrelated challenges: (1) spatial randomness, (2) limited computational power of individual EEDs, (3) device vulnerability, and (4) temporal randomness. Spatial randomness arises from the highly dynamic network topology, potentially leading to an insufficient number of EEDs in certain locations \cite{lit11}. Unlike conventional MEC or cloud computing, EEDs have limited computational resources, making parallel task execution across multiple devices essential to meet performance requirements. Device vulnerability presents another significant challenge, as these user-owned devices are subject to intermittent availability, uncertainty, and higher failure risks, thereby requiring explicit reliability considerations \cite{amer2023task}. Furthermore, temporal randomness emerges from fluctuations in offloading durations and task execution times, primarily due to the uncertainty associated with wireless channel conditions, signal-to-interference-plus-noise ratio (SINR) variability, task size diversity, and heterogeneous device capabilities. These intertwined challenges, including stochastic communication success and the temporal overlap between computation
and communication, highlight the critical need for a rigorous
spatiotemporal mathematical framework. Such a model is essential for accurately quantifying performance trade-offs in the EEC architecture, a gap that remains unaddressed in the existing literature.

In this paper, we quantify the interplay between communication and computation costs within large-scale millimeter-wave (mmWave) networks for EEC. Our primary contribution lies in developing the first spatiotemporal analysis for EEC, combining stochastic geometry (SG) and queueing theory, and uniquely employing an absorbing continuous-time Markov chain (ACTMC) to capture the dynamic interaction between task offloading via D2D communication and parallel computation across EEDs, which overlap in time.
The proposed system partitions computational tasks into smaller segments, which are offloaded to multiple EEDs to accelerate execution.

The proposed spatiotemporal analysis is readily applicable to data-parallel IoT workloads. Representative examples include distributed machine learning inference on sensor readings or image batches and real-time video analytics~\cite{wang2024paraloupe}, feature-map-partitioned deep-learning inference~\cite{zhao2018deepthings}, and privacy-aware task segmentation~\cite{razaq2021privacy}.
More recently, such latency- and reliability-critical execution has become equally important for emerging edge-enabled digital-twin services. For instance, digital-twin pipelines may require the dynamic deployment and continuous update of human digital twins across edge servers to support timely task execution under mobility and time-varying conditions~\cite{10540318}. Likewise, interactive digital-twin services impose tight round-trip communication and computation constraints to deliver responsive feedback and user experience under evolving twin states~\cite{11302884}. In such digital-twin pipelines, many update and interaction workloads are naturally partitionable and can be opportunistically executed in parallel over nearby EEDs to meet stringent timeliness and reliability requirements, precisely the regime quantified by our spatiotemporal EEC analysis.

To this end, we analytically evaluate the average task response delay, a fundamental performance metric in EEC that reflects its viability for supporting latency-sensitive applications such as data-parallel distributed learning and real-time processing. Specifically, we utilize SG to derive the offloading success probability, accounting for device locations, mmWave antenna characteristics, channel conditions, and network-wide interference. This probability determines the EED offloading rates utilized in our ACTMC model, which enables precise evaluation of the average task response delay.
In addition to delay, we consider the task completion probability as a metric to evaluate system reliability, an essential consideration in failure-prone EEC environments. This metric captures the likelihood that all task segments are successfully executed. Together, these two metrics provide a meaningful assessment of EEC system performance and guide informed decisions on task segmentation levels, balancing both latency and reliability requirements.

To summarize, this paper makes the following contributions:

\begin{itemize}
\item We propose the first spatiotemporal analytical framework for EEC that integrates SG with an ACTMC to jointly model D2D offloading and parallel computation in large-scale networks.
\item We derive tractable expressions for the average task response delay and task completion probability, characterizing the communication-computation trade-off that yields a delay-optimal task segmentation.
    \item We quantify the performance gains of location-aware EED selection and incorporate device failures, enabling explicit analysis of latency-reliability trade-off.
    \item We introduce a bias-based EEC-MEC collaboration scheme to mitigate system congestion in mmWave networks with limited line-of-sight (LoS) EED availability.
\item We validate the analysis via Monte Carlo simulations and sensitivity studies, yielding robust design guidelines across diverse network conditions.
\end{itemize}

The remainder of the paper is organized as follows. Section II reviews the related work. Section III introduces the baseline spatiotemporal model with random EED selection under abundant EED availability. Section IV extends the model to practical scenarios with limited EED availability, incorporating location-aware selection, device failures, and EEC-MEC collaboration. Section V presents the numerical and simulation results, and Section VI concludes the paper and discusses future work.

\section{Related Work}

Scheduling tasks in centralized cloud data centers has been extensively studied to improve the efficiency of parallel workflow execution across heterogeneous virtual machines (VMs). Complex applications are typically modeled as directed acyclic graphs (DAGs), where interdependent tasks are mapped to VMs and scheduled to optimize multiple objectives. Mohammadzadeh et al. investigated scientific workflow scheduling in green-cloud environments and proposed an improved chaotic binary Grey Wolf algorithm to minimize cost, makespan, and power consumption \cite{mohammadzadeh2021improved}. A
related study broadens the optimization scope to a system-level multi-objective
framework for enhanced resource efficiency and overall workflow performance in cloud
data centers \cite{mohammadzadeh2021energy}. Li et al.\ \cite{li2024makespan} further
extend workflow scheduling by incorporating security requirements, jointly optimizing
service cost and data protection when deploying workflows across cloud-based VMs.
Although effective for centralized infrastructures, these approaches are not designed
for the latency-sensitive and spatially distributed requirements of emerging 6G-enabled IoT and
AI-driven applications, motivating the shift toward MEC.

MEC has therefore been widely investigated as a means to offload computation from
resource-constrained devices to nearby edge servers. In \cite{jiang2022joint}, Jiang
et al.\ formulate a real-time optimization framework for joint task offloading and
resource allocation in MEC under a long-term energy constraint, where each task is
treated as an indivisible unit executed either locally or at the edge. Liu et al.\
\cite{lit23} develop an optimal stochastic computation offloading policy in a
single-user MEC system, allowing tasks to be processed locally, at the MEC server, or
in parallel across local and MEC processors. In \cite{chen2015efficient}, Chen et al.\
adopt a game-theoretic approach, designing an algorithm that converges to a Nash
equilibrium and achieves performance and scalability when many devices share MEC
resources. 

Queueing-aware MEC control has been studied for delay-sensitive services; for example, Yi et al.\ jointly design multi-user
computation offloading and uplink transmission scheduling with pricing-based incentives to mitigate congestion under
random task arrivals \cite{8606230}. Moreover, Cao and Cai \cite{8012473} formulate multichannel-contention cloudlet
offloading as a noncooperative game and propose a fully distributed learning algorithm that converges to a pure-strategy
Nash equilibrium without information exchange among users.
In more specialized scenarios, Moghaddasi et al.\ \cite{moghaddasi2024multi}
study vehicular MEC and propose a double deep Q-network (DDQN)-based multi-objective
offloading strategy that jointly optimizes latency, energy, and monetary cost while
integrating a blockchain layer for data integrity and coordination. Furthermore, Rahmani et
al.\ \cite{rahmani2025self} focus on IoT-MEC energy management, developing a
decentralized soft actor-critic framework for device-local, context-aware power control
with limited MEC coordination and reporting substantial energy and battery-life gains in
smart-home environments.

Beyond MEC, EEC further leverages nearby EEDs as cooperative workers for task execution. Azmy et al.\ develop an incentive-vacation queueing framework for reward-based EEC, where user-owned EEDs act as workers and incentive-coupled vacation queues with continuous-time Markov chains are used to characterize queueing delay and worker time 
in system, and to derive metrics for predicting worker participation dynamics under different incentive contracts \cite{azmy2024incentive}. Masoumi et al.\ apply EEC in industrial settings with mobile robots, using a hierarchical EEC/edge/cloud architecture with heuristic queue-aware scheduling and deadlock mitigation to coordinate movement and onboard processing tasks \cite{masoumi2025dynamic}. For digital twin services, El-Khatib et al.\ introduce a proactive scheme that maximizes a weighted service capacity objective by predicting EED resource usage and  forming collaborating worker groups to execute partitioned subtasks under deadline constraints \cite{el2025proactive}.

Learning-based orchestration has also been explored. Safavifar et al.\ propose a multi-objective deep reinforcement learning (DRL) workload orchestrator that assigns tasks to heterogeneous EEDs to reduce resource waste and energy consumption while maintaining a high task success rate \cite{safavifar2024multi}. Moreover, a DRL-based 
orchestrator for dependent composite tasks in EEC has been proposed in \cite{safavifar2025sustainable}, 
where applications are modeled as DAGs and decomposed into partitions that are offloaded 
to EEDs to minimize completion time and reduce MEC usage. From an experimental perspective, Drainakis et al.\  demonstrate service orchestration at the extreme edge over a 5G testbed, where an orchestrator manages containerized AI tasks on mobile and static EEDs via policy-based EED selection \cite{drainakis2025service}.

The works discussed above primarily focus on task offloading, resource allocation, and local dependability metrics under abstract network models, without explicitly capturing spatial randomness, network-wide interference, or the impact of node density and topology on performance. SG is therefore utilized, as in
\cite{lit11,elbayoumi2022edge,cheng2024design}, to capture the effect of network
geometry and interference in large-scale wireless systems with edge computing. In \cite{lit11}, task offloading in a mobile cloud computing network is analyzed under heterogeneous computational resources, and the network-wide outage probability is characterized. Elbayoumi et al.\ analyze edge computing in ultra-dense networks where
small cells equipped with edge computing servers form a Poisson point process (PPP),
and human-type users can associate with multiple small cells to partition and offload
elastic tasks that are processed in parallel at the edge and locally
\cite{elbayoumi2022edge}. More recently, Cheng et al.\ consider a MEC-aided uplink LoRa network with randomly distributed end devices modeled as a PPP and analyze the computation offloading success probability under interference and power control \cite{cheng2024design}.

Recent efforts have combined queueing theory with SG to jointly capture
network geometry and temporal dynamics, enabling a more complete spatiotemporal
characterization of large-scale wireless systems \cite{song2021age,9738803}. This
spatiotemporal viewpoint has motivated several MEC studies that jointly account for
communication and computation aspects of task execution.  In \cite{lit24}, a spatiotemporal model is proposed for large MEC networks,
where SG and queueing analysis are used to characterize both
communication and computation latency. In \cite{lit10}, the scalability of MEC-enabled wireless networks is
explored, and both communication and computation performance bounds are derived under a
variety of network parameters. However, each task is modeled as an indivisible job, and
a user can offload to at most a single MEC server. In \cite{lit17}, Gu et al.\ study a large-scale MEC
wireless network in which tasks can be computed locally or offloaded to a
MEC server, modeling the spatial distribution of access points (APs) and users via
SG and employing a two-dimensional discrete-time Markov chain to
capture the joint evolution of local-computation and offloading buffers and thereby
characterize end-to-end task execution performance.

Other works have extended this spatiotemporal modeling to incorporate dependability and
heterogeneous deployments. In \cite{lit13}, Emara et al.\ consider the joint impact of
network interference and parallel computing with multiple VMs residing on the same edge
server. Although computation is parallelized across VMs, each task is still offloaded to
a single centralized MEC server. In \cite{park2020mobile}, Park and Lee develop a spatiotemporal framework for
MEC-enabled heterogeneous networks, where multi-tier MEC servers and users are modeled
as PPPs and SG is combined with an M/G/1 queueing model to
characterize communication and computation latency.
 In \cite{gu2023communication}, Gu et al.\ develop a
spatiotemporal framework for MEC-enabled heterogeneous networks with
communication-computation-aware user association, where multi-tier MEC APs and users are
modeled as PPPs and SG is combined with queueing analysis to characterize the meta distribution of task offloading success
and the resulting latency.

The above works are summarized in Table~\ref{tab:related_work_summary}, which highlights
that existing spatiotemporal frameworks remain MEC-centric.  In these models, tasks are typically executed at infrastructure MEC servers, and users cannot leverage nearby EEDs as cooperative workers within the EEC paradigm. To the best
of our knowledge, this is the first work to develop a spatiotemporal analytical
framework for EEC, jointly
capturing the interplay between D2D connectivity, network-wide interference, and
parallel computing across EEDs to analytically characterize EEC performance.

\begin{table*}[t]
\centering
\caption{Summary of edge computing related works}
\label{tab:related_work_summary}
\footnotesize
\setlength{\tabcolsep}{4pt}
\begin{tabular}{lccccc}
\toprule
Work & MEC & EEC & Large-scale modeling (SG) & Spatiotemporal (SG+queueing) & Spatiotemporal EEC \\
\midrule
MEC task offloading and control  \cite{jiang2022joint,lit23,chen2015efficient,8606230,8012473,moghaddasi2024multi,rahmani2025self} & \checkmark & $\times$ & $\times$ & $\times$ & $\times$ \\
EEC workload orchestration \cite{azmy2024incentive,masoumi2025dynamic,el2025proactive,safavifar2024multi} & $\times$ & \checkmark & $\times$ & $\times$ & $\times$ \\
Hybrid MEC-EEC workload orchestration \cite{safavifar2025sustainable,drainakis2025service} & $\checkmark$ & \checkmark & $\times$ & $\times$ & $\times$ \\
SG-based works\cite{lit11,elbayoumi2022edge,cheng2024design} & \checkmark & $\times$ & \checkmark & $\times$ & $\times$ \\
Spatiotemporal-based works  \cite{lit24,lit10,lit17,lit13,park2020mobile,gu2023communication} & \checkmark & $\times$ & \checkmark & \checkmark & $\times$ \\
\textbf{This work} & \checkmark & \checkmark & \checkmark & \checkmark & \checkmark \\
\bottomrule
\end{tabular}
\end{table*}

\section{The Baseline Spatiotemporal Analysis}

This section presents a baseline spatiotemporal model, where EEDs are the only option that offers computational services. The EEDs are abundant, and their selection is made at random.
\subsection{Baseline System Model}\label{klkkl}

The computationally capable EEDs, also referred to as workers, are modeled via a PPP $\Phi \subset \mathbb{R}^2$ with intensity $\nu_w$. The EEDs offer their computational services to resource-constrained devices (e.g., IoT), which hereafter are referred to as requesters. The requesters are spatially distributed according to an independent PPP $\Omega \subset \mathbb{R}^2$ with intensity $\nu_r$. There is an  edge orchestrator that can be a BS or an AP, which organizes the offloading process between workers and requesters. In particular, the EEDs that have available computational power register their availability at the edge orchestrator, which in turn informs each requester about the availability of proximate EEDs. Specifically, when a requester decides to offload a task to the surrounding EEDs, it requests the edge orchestrator to assign available nearby EED resources. In that context, the edge orchestrator does not have the location information, so it sends the devices in a random order. It is assumed in this model that $\nu_w\gg \nu_r$, and hence, the edge orchestrator can readily allocate a unique worker to each task segment without contention. To utilize parallel computing and reduce response delay, the requester divides each computational task into $n$ smaller and equivalent segments to be offloaded and executed at different EEDs. Due to the heterogeneity of the computational powers of the EEDs, the execution time of each segment is exponentially distributed with mean $\frac{1}{n \mu_f}$, where $\mu_f$ is the task execution rate if computed at a single worker.

In compliance with 5G and beyond systems, the requesters utilize mmWave for D2D communications to offload segments to their proximate workers.  The high vulnerability of mmWave communications to blockage is considered via the general LoS ball blockage model~\cite{andrews2016modeling}. The devices within the distance of $R_L$ from the requester are considered LoS devices, and otherwise, any device located beyond that point is considered a non-line of sight (NLoS) device. Distance-dependent power-law path-loss is considered with exponents $\alpha_L$ and $\alpha_N$ for LoS and NLoS devices, respectively. All transmissions experience Nakagami multipath fading. Hence, the channel power gains have independent and identical gamma distribution parameters $N_L$ for LoS devices and $N_N$ for NLoS devices.

\begin{figure}[t!]
\centering
\includegraphics[width=180pt]{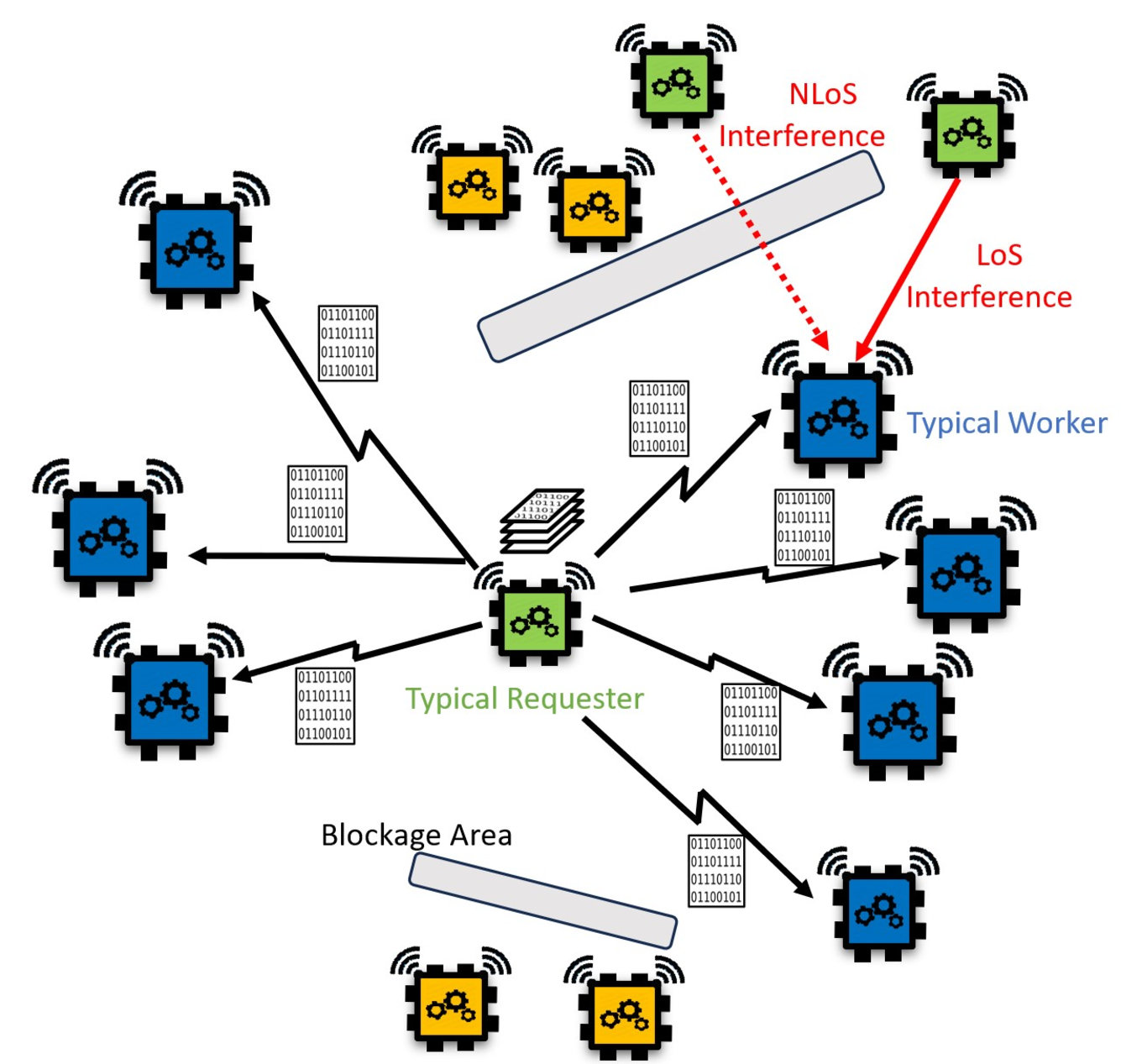}
\caption{The spatial system model: LoS workers (blue), NLoS workers (orange), and requesters (green). The typical requester offloads task segments to LoS workers. The typical worker receives the intended link from the typical requester, along with LoS and NLoS interference from other requesters.}
\label{fig_system_model}
\end{figure}

Universal frequency reuse and constant transmit power are utilized by all requesters. The requester and workers deploy antenna arrays for mmWave beamforming. The widely adopted sectored antenna model approximates the array patterns \cite{andrews2016modeling}. Accordingly, the main lobe gain is $M_{x}$, the side lobe gain is $m_{x}$, and the 3-dB beamwidth is $\theta_{x}$, where the subscript $x\in\{r,w\}$ is to differentiate between the antenna patterns of the requesters and workers. Without loss of generality, consider that a typical requester is located at the origin and can establish D2D links with proximate LoS EEDs only. Therefore, perfect antenna alignment is considered for the intended D2D link, and uniform random antenna alignment is considered for the interfering links. A pictorial illustration of the system model is shown in Fig.~\ref{fig_system_model}.

The requester is assumed to have one task divided into $n$ independent and equal segments.\footnote{Tasks with inter-segment dependencies, intermediate synchronization, or precedence constraints, as in DAG-structured workflows, fall outside the present scope of our model.}
 The segments are encapsulated into $n$ packets transmitted via D2D communications to different proximate workers. The workers are sequentially allocated since a single mmWave interface is available at the requester. The workers are selected randomly from the list of available LoS EEDs provided by the edge orchestrator.  Due to fluctuations in channel conditions, communication between the requester and the worker may encounter errors, which may require multiple attempts to deliver the segment and allocate the worker successfully.  Each segment transmission attempt via D2D communication takes $\tau_c$ seconds. The worker begins executing the segment immediately upon receiving the segment successfully. Upon receiving an acknowledgment (ACK) indicating a successful transmission, the requester offloads the remaining segments to other available LoS workers. Conversely, if a negative acknowledgment (NACK) is received, it indicates a transmission failure, prompting the requester to retransmit the same task segment until successful delivery. The ACK and NACK notifications are assumed to be transmitted over a perfect feedback channel. 
The result of each worker’s assigned segment is returned to the requester as soon as that worker finishes executing it. We assume the communication time for returning these segment results is negligible, which aligns with our target domain of IoT monitoring and distributed inference pipelines. In such
scenarios, outputs such as classification labels or control commands are compact relative to input data like sensor streams or
high-resolution images. This simplification is also consistent with standard practice in related literature \cite{jiang2022joint,cheng2024design,gu2023communication}.

\subsection{Offloading Success Probability with Random EED Selection}

To calculate the average task response delay, we first need to obtain the average segment offloading time. The worker correctly receives the segment if the SINR is above a given threshold $\xi$. Otherwise, the segment has to be retransmitted. Hence, the first step in investigating the response delay is to find the D2D communication success probability between the requester and the randomly selected LoS worker. Such probability will be utilized later within an ACTMC to find the average task response delay.  
Following ~\cite{andrews2016modeling}, the received SINR at the intended worker is given by
\begin{equation}
\label{sinr}
    {\rm SINR} = \frac{h_0 M_r M_w C_L r_0^{-\alpha_L}}{\sigma^2+ I_N+I_L}, 
\end{equation}
and the successful D2D transmission probability  of a segment can be expressed as
\begin{equation}\label{eq_sinr}
    p_s= \mathbb{P}\left\{\text{SINR}>\xi \right\} =  \mathbb{P}\left\{ \frac{h_0 M_r M_w C_L r_0^{-\alpha_L}}{\sigma^2+ I_N+I_L}  >\xi \right\}.
\end{equation}
Here, $h_0$ is the intended channel power gain, $C_L$ is the intercept of the LoS channel, $r_0$ is the distance between the requester and the intended LoS worker, $I_{L}$ is the aggregate interference from other active LoS requesters, $I_{N}$ is the aggregate interference from other active NLoS requesters. Moreover, $\sigma^2$ denotes the normalized noise power, i.e.,
$\sigma^2 \triangleq \sigma_{\mathrm{th}}^{2}/P_t$, where $\sigma_{\mathrm{th}}^{2}$
is the thermal noise power over bandwidth $B$ and $P_t$ is the transmit power
of each active requester. 
 Let $\Omega_L \subset \Omega$ and $\Omega_N = \Omega \setminus \{(\Omega_L) \cup (0,0)\}$ be the point processes of the LoS and NLoS requesters, respectively. Then, the LoS and NLoS interference terms as described in \cite{andrews2016modeling}, are then expressed by 
\begin{equation}\label{eqIagg1}
        I_L = \sum_{i>0:\bf{x}_i\in\Omega_L} h_i D_i C_L \left\|\bf{x}_i\right\|^{-\alpha_L}, 
\end{equation}
and
\begin{equation}\label{eqIagg2}
        I_N = \sum_{i>0:\bf{y}_i\in\Omega_N} g_i D_i C_N \left\|\bf{y}_i\right\|^{-\alpha_N},
\end{equation}
where $h_i$ is the $i^{th}$ LoS interfering link channel power gain, $g_i$ is the $i^{th}$ NLoS interfering link channel power gain, $C_N$ is the intercept of the NLoS channel, $\left\|\cdot\right\|$ is the Euclidean norm, and $D_i$ is the antenna gain for the $i^{th}$ interfering requester in $\Omega_L$ or $\Omega_N$. Given the sectored antenna model and the uniformly random alignment between a typical worker and an interfering requester, $D_i$ is a discrete random variable with four possible outcomes, each corresponding to a specific antenna gain scenario. These scenarios reflect the four possible alignments between the worker and the interferer requester, each with a specific probability. The distribution is $\mathbb{P}\{D_i = a_k\} = b_k$ for $k \in \{1, 2, 3, 4\}$, with $a_k$ and $b_k$ as defined in Table \ref{directivity_gain}.

The D2D transmission success probability given by  (\ref{eq_sinr}) is characterized in Theorem 1.
\begin{theorem}
The spatially averaged probability of successful segment offloading via mmWave D2D communication to a randomly selected LoS worker from $\Phi_w$ is given by
\begin{equation}
p_s = \int_{0}^{R_L}\sum_{n=1}^{N_L} {N_L \choose n}
 \frac{2 r_0 (-1)^{n+1} e^{M_n(\xi)\sigma^2 - W_n(\xi) - Z_n(\xi)}}{R_L^2}dr_0,
\label{eqCoverageAll}
\end{equation}
where $ M_n(\xi) = - \frac{\eta_L n r_0^{\alpha_L} \xi}{C_L M_r M_w}$, while $W_n(\xi) $ and $ Z_n(\xi)$ are given by
\begin{equation}
  W_n(\xi) = 2 \pi \nu_r  \sum_{k=1}^4 b_k \int_{0}^{R_L}
    \Bigg( 
    1 - \frac{1} {\Big(
    1 + \frac{\eta_L \bar{a}_k n \xi (\frac{r_0}{x})^{\alpha_L}}{N_L}
    \Big)^{N_L}}
    \Bigg)x dx,
    \label{eqCoverageW}
\end{equation}
\begin{equation}
  Z_n(\xi)= 2 \pi \nu_r  \sum_{k=1}^4 b_k \int_{R_L}^{\infty}
\Bigg( 
1 - \frac{1}{\Big(
1 + \frac{n_L \bar{a}_k n \xi C_N r_0^{\alpha_L}}{C_L x^{\alpha_N} N_N}
\Big)^{N_N}}
\Bigg) x dx.
\label{eqCoverageZ}
\end{equation}
Here, $\bar{a}_k = \frac{a_k}{M_r M_w}$, and $b_k$ along with $a_k$ for $1\leq k\leq 4$ are defined in Table \ref{directivity_gain}.
\begin{IEEEproof}
The proof can be found in Appendix A
\end{IEEEproof}
\end{theorem}
\begin{table}[!t]
\renewcommand{\arraystretch}{1.3}
\caption{Directivity Gain and Probability}
\label{directivity_gain}
\centering
 \begin{tabular}{|c|c|c|}
\hline
$k$ & \textbf{\( a_k \)} & \textbf{\( b_k \)} \\
\hline
1 & \( M_w M_r \) & \( \frac{\theta_w}{2\pi} \frac{\theta_r}{2\pi} \) \\
\hline
2 & \( M_w m_r \) & \( \frac{\theta_w}{2\pi} \left( 1 - \frac{\theta_r}{2\pi} \right) \) \\
\hline
3 & \( m_w M_r \) & \( \left( 1 - \frac{\theta_w}{2\pi} \right) \frac{\theta_r}{2\pi} \) \\
\hline
4 & \( m_w m_r \) & \( \left( 1 - \frac{\theta_w}{2\pi} \right) \left( 1 - \frac{\theta_r}{2\pi} \right) \) \\
\hline
\end{tabular}
\end{table}

\subsection{Average Task Response Delay Calculation}
The task response delay is defined as the time needed to process the $n$ segments, beginning when the requester starts offloading the first segment to the allocated EED and concluding when all $n$ segments have been executed. This delay encompasses both communication and computation components, along with their interactions. However, due to the randomness in factors such as EED locations and availability, channel gains, computational power, and failure rates, the delay calculated in this study is represented as an average measure, referred to as the average task response delay.
 Note that the average time that one segment takes to be executed is $ T_o = \tau_h + \tau_f$, where $\tau_h$ is the average offloading time that the requester takes to offload a segment to a randomly selected EED, and $\tau_f$ is the average time the segment takes to be executed at the intended EED. In this context, $\tau_h = \tau_c / p_s $, where $p_s$ is the probability given in Theorem~1, and $\tau_c$ is the average D2D communication time in mmWave networks.

The average task response delay cannot be simply represented as the sum of individual segment delays ($\neq n T_o$), as this would ignore both the parallel processing within the system and the overlap between communication and computation times. The system’s complexity, influenced by the interactions between simultaneous segment processing, the stochastic nature of communication offloading, and the overlapping of communication and computation times, combined with the fact that allocation and completion events can happen at any moment, requires modeling using an ACTMC.
To this end,  the successful offloading probability estimated in Theorem~1 is a core building block of the ACTMC, where the average offloading rate $\lambda_h = 1/\tau_h$.
Next, we delve into the foundational ACTMC and embedded discrete-time Markov chain (EDTMC) employed.

\subsubsection{ACTMC and EDTMC}
The state set of the ACTMC is represented as $\mathcal{S} = \{ {\bf z}=(x_f , x_c) \mid \sum_j x_j \leq n ; j \in \{f,c\} \}$, where $x_f \in \{0,1,2,\cdots, n\}$ denotes the number of workers that have finished their assigned segments successfully, and $x_c \in \{0,1,2,\cdots, n\}$ denotes the number of workers that are executing the assigned segments.
For each task, ACTMC starts at the state ${\bf z}_1= (0,0)$, where the requester has a task that is sliced to $n$ segments but has not yet allocated any worker. Each time the requester succeeds in allocating a LoS EED via mmWave D2D transmission, a transition occurs from the current state ${\bf z}_i= (x_f,x_c)$ to the next state ${\bf z}_j= (x_f,x_c+1)$. Moreover, each time a worker is retired because of segment completion, a transition from state  ${\bf z}_i= (x_f,x_c)$ to ${\bf z}_j= (x_f+1,x_c-1)$ occurs.  Since the requester needs only $n$ workers, then $x_c+x_f \leq n$ and ${\bf z}_L =(n,0)$ is the absorbing state that implies the termination of the ACTMC, where $L$ is the total number of states in the system.

Following the criterion mentioned above, segments offloading and execution at the EEDs can be tracked with an ACTMC with the following two-level hierarchical generator matrix
\begin{equation}
 \bf{Q} = \!\!\!\!\!\kbordermatrix{
  x_f & 0 & 1 & 2 & 3 & \cdots & n\\
    0 & \bf{K}_0 & \bf{H}_{0,1} &\bf{0} & \bf{0} & \cdots & \bf{0} \\
  1 &\bf{0} & \bf{K}_1 & \bf{H}_{1,2} & \bf{0} & \ddots & \bf{0}\\
  2 &\bf{0} & \bf{0} & \bf{K}_2 &\bf{H}_{2,3} &  \ddots & \bf{0} \\
 \vdots & \vdots &  \ddots & \ddots & \ddots & \ddots & \vdots \\
  n-1 & \bf{0} & \cdots & \bf{0} & \bf{0}  & \bf{K}_{n-1} & \bf{H}_{n-1,n}  \\
  n & \bf{0} &  \cdots & \bf{0} & \bf{0}  & \bf{0} & \bf{0} \\
  },
\end{equation}
where ${\bf Q}$ is a block matrix of size $(n+1)\times(n+1)$ that tracks the number of finished workers $x_f$. Since the task is finished upon the completion of the $n$ segments, then the state $x_f=n$ is the absorbing state that indicates the termination of the edge computing.  Within each level of $\bf{Q}$, the sub-matrices ${\bf K}_m$ and ${\bf H}_{m,m+1}$ track the number of allocated workers $x_c$.\footnote{The hierarchical structure of the proposed ACTMC enables scalable analysis and eliminates the need to visualize the full $(n+1) \times (n+1)$ generator matrix, whose entries are submatrices.} Exploiting the fact that $x_c+x_f \leq n$, the matrix ${\bf H}_{m,m+1}$ is of size $(n-m)\times (n-m-1)$ that tracks $x_c$ due to the completion of a segment by any of the workers. Let ${\bf {H}}_{m,m+1}(i,j)$, with $i\in \{0,1,2,\cdots,n-m\}$ and $j\in \{0,1,2,\cdots,n-m-1\}$, denote the $(i,j)^{\text{th}}$ element of the matrix ${\bf {H}}_{m,m+1}$. Then, due to the parallelism in the computing at the EEDs along with the fact that only one worker can finish at a given instance, ${\bf H}_{m,m+1}$ is given by 
\begin{equation}\label{ghgg}
\mathbf{H}_{m,m+1}(i,j)=
\begin{cases}
i\,\mu_f, & i=j+1,\\
0, & \text{otherwise}.
\end{cases}\,
\end{equation}

Using a similar argument, the matrix ${\bf {K}}_m$ is of size $(n-m+1) \times (n-m+1)$ that tracks $x_c$ upon allocating  new workers. Let ${\bf {K}}_m(i,j)$, with $i,j\in \{0,1,2,\cdots,n-m\}$ denote the $(i,j)^{\text{th}}$ element of the matrix ${\bf {K}}_m$. Accordingly, due to the sequential worker allocation, ${ {\bf K}}_m(i,j)$ is given by
\begin{equation}\label{hgsdj}
\mathbf{K}_m(i,j)=
\begin{cases}
-(\lambda_h + i\mu_f), & i=j \text{ and } i<n-m, \\
\lambda_h,              & i=j-1 \text{ and } i<n-m, \\
-(n-m)\mu_f,            & i=j=n-m, \\
0,                      & \text{otherwise},
\end{cases}
\end{equation}
where $\lambda_h = p_s / \tau_c$ is the offloading rate, $p_s$ is the D2D transmission success probability given in (\ref{eqCoverageAll}), and $\tau_c$ is the time required for each D2D transmission attempt.

The average task response delay cannot be directly obtained for the matrix ${\bf Q}$. Instead, we first need to obtain the EDTMC of ${\bf Q}$ and the average sojourn time at each state. The EDTMC of ${\bf Q}$ is given by 
\begin{equation}\label{gdshs}
 \bf{P} = \!\!\!\!\!\kbordermatrix{
  x_f & 0 & 1 & 2 & 3 & \cdots & n\\
    0& \bf{\mathcal{K}}_0 & \bf{\mathcal{H}}_{0,1} &\bf{0} & \bf{0} & \cdots & \bf{0} \\
    1 &\bf{0} & \bf{\mathcal{K}}_1 & \bf{\mathcal{H}}_{1,2} & \bf{0} & \ddots & \bf{0}\\
    2 &\bf{0} & \bf{0} & \bf{\mathcal{K}}_2 &\bf{\mathcal{H}}_{2,3} & \ddots & \bf{0} \\
    \vdots & \vdots & \ddots  & \ddots & \ddots & \ddots & \vdots \\
    n-1 & \bf{0} & \cdots &  \bf{0} & \bf{0}  & \bf{\mathcal{K}}_{n-1} & \bf{\mathcal{H}}_{n-1,n}  \\
    n & \bf{0} &  \cdots & \bf{0} & \bf{0}  &\bf{0} & 1 \\
  },
\end{equation}
where $\mathcal{K}_m$ and ${\bf \mathcal{H}}_{m,m+1}$ track the transition probabilities due to worker allocation and segment completion, respectively. Moreover, the matrices $\mathcal{K}_m$ and $\mathcal{H}_{m,m+1}$ are given by
\begin{equation}\label{sbddsd}
\mathcal{K}_m(i,j)=
\begin{cases}
\dfrac{\lambda_h}{\lambda_h + i\mu_f}, & i=j-1,\\
0,                                      & \text{otherwise},
\end{cases}
\end{equation}
\begin{equation}\label{gsfhajkaSA}
\mathcal{H}_{m,m+1}(i,j)=
\begin{cases}
\dfrac{i\,\mu_f}{\lambda_h + i\mu_f}, & i=j+1 \text{ and } i<n-m,\\
1,                                    & i=n-m,~j=n-m-1,\\
0,                                    & \text{otherwise}.
\end{cases}
\end{equation}


\subsubsection{Average Time until Absorption}

After formulating the ACTMC and obtaining the matrices {\bf Q} and {\bf P}, we now utilize those to calculate the average task response delay. The ACTMC has an absorbing state that is reached once all $n$ segments have been successfully executed at the allocated EEDs. Based on that, the average task response delay is equivalent to the average time until absorption in that state. To calculate the average time until absorption, let $x_{c_i}\in {\bf z}_i$ be the number of allocated workers in state $ {\bf z}_i$, then the average sojourn time $t_{{\bf z}_i,{\bf z}_j}$ is given by 
\begin{equation}
t_{{\bf z}_i,{\bf z}_j}=\left\{\begin{matrix} 
\frac{1}{x_{c_i} \mu_f},   &  \begin{matrix}\text{if the transition from ${\bf z}_i$ to ${\bf z}_j$   } \\ \text{ is due to segment completion, }\end{matrix}  \\
& \\
\frac{1}{\lambda_h}, & \begin{matrix}\text{if the transition   from ${\bf z}_i$ to ${\bf z}_j$   } \\ \text{ is due to worker allocation.} \end{matrix}
\end{matrix}\right.
\label{t_z_values}
\end{equation}

Equipped with ${\bf P}$ and $t_{{\bf z}_i,{\bf z}_j}$, the average task response delay is given in Theorem 2.
\begin{theorem}
The average task response delay in the extreme edge computing networks with mmWave D2D communications and $n$ randomly allocated  workers is given by 
\begin{equation}
    T_A = \boldsymbol{\alpha} ({\bf I} - {\bf P}_T)^{-1} {\bf w},
    \label{time_equation}
\end{equation}
where $\boldsymbol{\alpha} = [1, 0, 0, \dots, 0]$ with a dimension of $1 \times L$ represents the system's initial state,
 ${\bf I}$ is the identity matrix, ${\bf P}_T$ is the transition probability of the transient states only in ${\bf P}$, which is obtained by excluding the transitions to the absorbing state (the last row and column of ${\bf P}$). The column vector ${\bf w}$ contains the average sojourn times at states ${\bf z}_i$, which are given by $w_{{\bf z}_i} = \sum_{{\bf z}_j} {\bf P}({\bf z}_i, {\bf z}_j) t_{{\bf z}_i, {\bf z}_j}$, where ${\bf P}({{\bf z}_i}, {{\bf z}_j})$ is the transition probability from state ${{\bf z}_i}$ to ${{\bf z}_j}$.\footnote{In line with the hierarchical structure of ${\bf P}$, we use two-dimensional indexing for its elements. Specifically, $\mathbf{P}\big({{\bf z}_i},{{\bf z}_j}\big)= \mathbf{P}\big((x_{f_i},x_{c_i}),(x_{f_j},x_{c_j})\big)$ is the $(x_{c_i}, x_{c_j})$ element of the $(x_{f_i}, x_{f_j})$ sub-matrix in $\mathbf{P}$.}
\begin{IEEEproof}
The proof can be found in Appendix B
\end{IEEEproof}
\end{theorem}

\section{Advanced Spatiotemporal Analysis}

To address the limitations of the baseline model, we introduce an advanced model that analyzes location-aware EED selection and accounts for potential EED failures. In addition, we propose task completion probability as a reliability metric, particularly relevant in scenarios with limited and/or failure-prone workers. This metric enables the quantification of system robustness under uncertainty and further enriches the performance evaluation of EEC environments by incorporating reliability in addition to latency. Furthermore, we investigate the impact of limited EED intensity relative to requester intensity on EEC performance and examine a bias model that enables collaboration between EEC and MEC, aiming to enhance system performance, particularly when contention over available LoS EEDs is high.

\subsection{Advanced System Model}

The advanced system model still shares some similarities with the baseline model described in Section \ref{klkkl}. Such similarities include the fact that workers and requesters are still modeled as PPPs with intensities $\nu_w$ and $\nu_r$, respectively. In addition, the requester can only allocate the LoS EEDs due to 
blockages. Thus, offloading is performed after obtaining information on surrounding LoS EEDs from the edge orchestrator, which maintains EED availability information.
Unlike the baseline model that assumes no scarcity of available EED, the advanced system model considers a limited number of available EEDs for each requester. In this case, when a requester decides to offload a task, the orchestrator maintains a pool with a limited number of EEDs, where the average number of available EEDs is  $\nu_w \pi R_L^2$.  

Moreover, when a requester probes the orchestrator for information about surrounding LoS EEDs, their locations are also included in the provided EED information.
As a result, the offloading shifts from randomly selecting a device to preferring the closest $i^{th}$ device for the $i^{th}$ offloading action. This refined approach aims to improve the probability of successful offloading. By selecting a nearby device, the signal quality improves, leading to a decrease in path loss and an overall increase in both the successful offloading probability and the offloading rate.

In addition, effectively handling EED failures is essential for enabling realistic EEC operations; therefore, failure events are explicitly considered. Specifically, if an EED fails during task execution, the requester allocates a replacement EED. It is important to note that the execution time impacts the failure likelihood: the longer a device operates to complete a task, the more exposed it is to disruptions and dropouts.  To quantify this, we define the failure rate as $\gamma = \frac{\mu_f}{l}$, where $l$ represents the system reliability parameter, indicating that an EED, on average, fails $l$ times less frequently than it successfully executes a task. For a task divided into $n$ segments, the failure rate for each device is expressed as $\gamma_n = \frac{\gamma}{n}$.

Finally, to address congestion in practical scenarios resulting from multiple requesters competing for the limited available LoS EEDs, a collaborative offloading approach involving both EEC and MEC is explored to reduce the average response delay. This method considers a bias factor, denoted by $\alpha$, which represents the proportion of requesters offloading their tasks to EEDs.

\subsection{Distance-based Successful Offloading  Probability}\label{lo_aw_offla}

Since EED allocation is done by selecting the closest device to the requester. Let ${\bf R} = \{R_{(1)}, R_{(2)}, ..., R_{(k)}, ..., R_{(n)}\}$ be the sorted distance vector of all the LoS EEDs, where $k$ represents the rank of the EED in the sorted vector, such that $R_{(1)} = \bf min\{R\}$ and $R_{(n)} = \bf max\{R\}$.  Theorem 3 represents the segment's successful offloading probability when selecting a new device based on its rank.

\begin{theorem}
The spatially averaged probability of successful segment offloading via mmWave D2D communications for a distance-based selected LoS worker is given by
\footnotesize
\begin{equation}
p_{s_{(k)}} = \int_{0}^{R_L}\sum_{n=1}^{N_L} {N_L \choose n}
 (-1)^{n+1} e^{M_n(\xi)\sigma^2 - W_n(\xi) - Z_n(\xi)}f_{(k)}(r_0)dr_0\, ,
\label{coverage_order}
\end{equation}
\normalsize
where $ M_n(\xi) = - \frac{\eta_L n r_0^{\alpha_L} \xi}{C_L M_r M_w}$, $W_n(\xi) $ and  $ Z_n(\xi)$ are given in  (\ref{eqCoverageW}) and (\ref{eqCoverageZ}), $f(x) = 2r_0/R_L^2$, $F(x) = r_0^2/R_L^2$, $V = \pi \nu_w R_L^2$, and $f_{(k)}(x)$ is given by  
\begin{equation}
      f_{(k)}(x) = \frac{V^{k} e^{-V} f(x) F(x)^{k-1}}{(k-1)!} e^{-V [F(x) - 1]}.
      \label{ORDERED_PDF}
\end{equation}
\begin{IEEEproof}
The proof can be found in Appendix C
\end{IEEEproof}
\end{theorem}

The offloading probability $p_{s_{(k)}}$ requires changing the ACTMC  and EDTMC to be level-dependent, which means that any selected device has its own offloading rate. This offloading rate is represented as $\lambda_{h_k} = p_{s_k}/\tau_c$, where $p_{s_k}$ is the successful offloading  probability of the $k^{th}$ closest EED to the requester. The updated matrices are given as follows: 

\begin{equation}
\mathbf{K}_m(i,j)=
\begin{cases}
-(\lambda_{h_{i+1}} + i\mu_f), & i=j \text{ and } i<n-m,\\
\lambda_{h_{i+1}},             & i=j-1 \text{ and } i<n-m,\\
-(n-m)\mu_f,                   & i=j=n-m,\\
0,                             & \text{otherwise},
\end{cases}
\end{equation}
\begin{equation}
\mathcal{K}_m(i,j)=
\begin{cases}
\dfrac{\lambda_{h_{i+1}}}{\lambda_{h_{i+1}} + i\mu_f}, & i=j-1,\\
0,                                                     & \text{otherwise},
\end{cases}
\end{equation}
and
\small
\begin{equation}
\mathcal{H}_{m,m+1}(i,j)=
\begin{cases}
\dfrac{i\,\mu_f}{\lambda_{h_{i+1}} + i\mu_f}, & i=j+1 \text{ and } i<n-m,\\
1,                                            & i=n-m,\; j=n-m-1,\\
0,                                            & \text{otherwise}.
\end{cases}
\end{equation}
\normalsize

The practical implementation of distance-based EED selection relies on efficient location acquisition mechanisms with minimal overhead. The edge orchestrator can obtain EED location information through several methods native to modern wireless systems. The orchestrator can leverage existing uplink reference signals, already transmitted by EEDs for standard communication functions like channel state information (CSI) estimation, to determine EED locations~\cite{yang2024positioning}. This approach requires no dedicated additional signaling from the EEDs, shifting the computational burden entirely to the orchestrator~\cite{yang2024positioning}. Alternatively, EEDs can self-report their positions using onboard GNSS/GPS capabilities by appending location data to the registration messages they send to the orchestrator to indicate their availability~\cite{musa2015trading}. This method introduces only a few bytes of overhead and operates at a low frequency relative to task offloading cycles, ensuring the signaling cost remains minimal~\cite{musa2015trading}. Looking forward, emerging 6G Integrated Sensing and Communication (ISAC) paradigms are expected to further streamline this process, as high-precision device location can be inferred directly by analyzing communication signals and their interactions with the environment, thereby eliminating dedicated positioning signaling and its associated overhead~\cite{11270002}.

\subsection{Modeling EEDs Failure}
To reflect the changes in the proposed model after introducing the system reliability parameter, the matrices $K_m$ and $H_m$ are now represented as follows:
\begin{equation}
K_m(i,j)=
\begin{cases}
i\,\gamma_n,                               & j=i-1,\\
-i(\gamma_n+\mu_f)-\lambda_{h_{i+1}},      & i=j,\; i<n-m,\\
-(n-m)(\gamma_n+\mu_f),                    & i=j=n-m,\\
\lambda_{h_{i+1}},                         & j=i+1,\; i<n-m,\\
0,                                         & \text{otherwise},
\end{cases}
\end{equation}
\begin{equation}
H_m(i,j)=
\begin{cases}
i\,\mu_f, & j=i-1,\\
0,        & \text{otherwise}.
\end{cases}
\end{equation}
Consequently, ${\bf \mathcal{K}}_m(i,j)$ and ${\bf \mathcal{H}}_{m,m+1}(i,j)$ are modified as follows:
\small
\begin{equation}
\mathcal{K}_m(i,j)=
\begin{cases}
\dfrac{i\,\gamma_n}{i(\gamma_n+\mu_f)+\lambda_{h_{i+1}}}, & j=i-1,\; i<n-m,\\
\dfrac{\gamma_n}{\gamma_n+\mu_f},                        & i=n-m,\; j=n-m-1,\\
\dfrac{\lambda_{h_{i+1}}}{i(\gamma_n+\mu_f)+\lambda_{h_{i+1}}}, & j=i+1,\; i<n-m,\\
0,                                                       & \text{otherwise},
\end{cases}
\end{equation}
\normalsize
\footnotesize
\begin{equation}
\mathcal{H}_{m,m+1}(i,j)=
\begin{cases}
\dfrac{i\,\mu_f}{i(\gamma_n+\mu_f)+\lambda_{h_{i+1}}}, & j=i-1,\; i<n-m,\\
\dfrac{\mu_f}{\gamma_n+\mu_f},                         & i=n-m,\; j=n-m-1,\\
0,                                                     & \text{otherwise}.
\end{cases}
\end{equation}
\normalsize
Moreover, the average sojourn time 
$t_{z_i, z_j} = \frac{1}{x_{ci} \gamma_n}$  if the transition from  $z_i$ to  $z_j$ is due to a failure.

\subsection{Task Completion Probability}

To further assess the resilience of the EEC system under realistic conditions, we introduce the task completion probability metric. It quantifies the likelihood that all parallelized task segments are successfully executed by dynamically recruited EEDs under conditions of limited worker availability and/or potential device failures.
While the average task response delay serves as the primary metric for evaluating latency performance and identifying the optimal number of task segments \( n \), it does not guarantee successful task completion. In contrast, task completion probability offers insight into the reliability of the system by capturing the probability that all task segments are completed successfully.
By analyzing both metrics jointly, we can determine the segmentation level \( n \) that satisfies not only delay minimization but also a desired reliability threshold. This enables informed decision-making for applications with varying priorities. For instance, safety-critical systems may prioritize reliability, whereas latency-sensitive tasks, such as real-time video processing, may focus on minimizing delay.

Mathematically, let \(\rho_t(i)\) denote the probability that all \( n \) task segments are successfully completed, starting from system state \( i \). This probability is computed recursively as:
\begin{equation}
\rho_t(i) = 
\begin{cases} 
1, & \text{if } i \text{ is a success state}, \\
0, & \text{if } i \text{ is a failure state}, \\
\sum_j P(i,j) \cdot \rho_t(j), & \text{otherwise (transient state)},
\end{cases}
\label{eq:rho}
\end{equation}
where \( P(i,j) \) is the transition probability from state \( i \) to state \( j \), and \( j \) indexes the successor state. A \textit{success state} occurs when all \( n \) segments have been executed, whereas a \textit{failure state} corresponds to scenarios in which the system exhausts its capacity to recover due to persistent failures and/or low worker intensity.
The system-wide task completion probability, denoted by \( \rho_t \), is defined as \( \rho_t(0) \), where \( i = 0 \) corresponds to the initial state with no workers recruited and no segments completed. The transition probabilities \( P(i,j) \) are derived from the failure-aware EDTMC, capturing the effects of failure events and recovery dynamics.

\subsection{Worker Status and EED Bias Factor}
Let $\alpha \in [0,1]$ denote the bias factor, representing the fraction of requesters that offload to EEDs. This yields a requester intensity of $\nu_{r_{\alpha}} = \alpha \nu_r$ for EEC offloading, while the remaining $(1-\alpha)\nu_r$ offload to MEC. This approach enables studying a combined EEC-MEC offloading strategy that balances computational load between both resources.

The availability of a worker (EED) depends on the following parameters: 
\begin{enumerate}
    \item The intensity of other workers $\nu_w$, since the probability of an EED being available decreases with fewer workers.
    \item The task execution rate $\mu_f$, since the higher the execution rate, the higher the probability that the EED will be idle.
    \item The intensity of the requesters $\nu_r$, since each requester needs to allocate EEDs to execute its task, and thus the higher the number of requesters, the higher the probability that the EED will be busy executing a task.
\end{enumerate}

\begin{figure}[h!]
    \centering
        \begin {tikzpicture}[-latex ,auto ,node distance =4 cm and 5cm ,on grid ,
        semithick,
        state/.style ={ circle ,top color = white, draw , text=black , minimum width = 1 cm}]
        \node[state] (C)
        {$busy$};
        \node[state] (A) [left=of C] {$idle$};
        \path (C) edge [bend left =25] node[below =0.15 cm] {$n\mu_f$} (A);
        \path (A) edge [bend right = -25] node {$\frac{n\nu_r}{\nu_w}$} (C);
        \end{tikzpicture}
    \caption{Worker CTMC}
    \label{fig:workers_CTMC}
\end{figure}
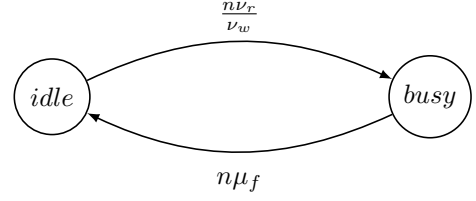

Consequently, the status of each worker is represented by a continuous-time Markov chain (CTMC), where the worker can be in one of two states: $idle$, indicating it has no current task segment and is ready to receive one, and $busy$, indicating it is actively computing a segment. Fig.~\ref{fig:workers_CTMC} illustrates the states and transitions in the worker's CTMC.

To model congestion, this CTMC determines the intensity of idle EEDs in scenarios where requesters compete for available EEDs. Let ${\bf \pi} = \{ \pi_{idle}, \pi_{busy} \}$ be the vector that represents the probability that an EED can be at the $idle$ state or the $busy$ state, respectively. The value of ${\bf \pi}$ is obtained by solving $\bf \pi Q = 0$ and $\sum_{s}\pi_s = 1$ where $s \in \{idle, busy\}$ and $Q$ is the state transition matrix of the CTMC. After solving, the value of $\pi_{idle}$, which reflects the probability of an EED being idle, is given by
\begin{equation}\label{lkjhdsg}
    \pi_{idle} = \frac{\mu_f}{\mu_f + \frac{\nu_r}{\nu_w}}.
\end{equation}
 Note that the steady state solution $\pi_{idle}$ does not depend on the number of task segments $n$. This is because, as the number of task segments increases, the probability of an EED being assigned a task and becoming busy also increases. However, the EED will complete its task more rapidly, returning to an idle state again.

Utilizing $\pi_{idle}$, the intensity of the EEDs that are idle is $\nu_{w_{idle}} = \pi_{idle}*\nu_w$, and the intensity of the EEDs that are busy is $\nu_{w_{busy}} = \nu_w - \nu_{w_{idle}}$.
As explained before, the value of $\nu_{w_{idle}}$ depends on the value of $\nu_r$, so low values of the bias factor $\alpha$ can be utilized to reduce the intensity of the requesters that will offload to EEDs, which will increase the number of available EEDs. Conversely, decreasing the value of $\alpha$ will increase the load on the MEC. To achieve the balance in the average response delay in both EEDs and MEC, let $\tau_{\alpha} = \alpha \times \tau_{EEDs} + (1-\alpha) \times \tau_{MEC}$ be the average response delay for the EEDs and the MEC, and the optimal value of $\alpha$ will be the one that results in the lowest value of $\tau_{\alpha}$.

It is worth highlighting that by congestion we mean scenarios where requesters face difficulty in finding available EEDs due to system wide resource scarcity. This does not refer to multiple requesters competing for specific idle EEDs, because the edge orchestrator inherently prevents such device level contention through centralized management of EED assignments. The orchestrator maintains exclusive control over worker allocation, ensuring that each available EED is assigned to at most one requester at any time.
Instead, congestion manifests mathematically through the effective idle worker intensity $\nu_{w_{\text{idle}}}$, which serves as a fundamental parameter in our spatiotemporal analysis and directly quantifies the system wide availability of computational resources. When $\nu_{w_{\text{idle}}}$ is low, indicating scarce EED availability, the allocation process becomes more challenging: requesters are forced to connect to more distant EEDs. This degradation reduces the D2D success probability, increases retransmission rates, and prolongs worker allocation time, thereby increasing the task response delay. To mitigate these effects, we tune the bias factor $\alpha$ to control the EEC-MEC split under congestion.

\section{Numerical Results and  Simulation}

\begin{table}[!t]
\caption{Numerical Parameters}
\label{table_params}
\centering
\footnotesize
\renewcommand{\arraystretch}{1.1}
\setlength{\tabcolsep}{4pt}
\resizebox{0.95\columnwidth}{!}{%
\begin{tabular}{|l|l|}
\hline
Parameter & Value\\
\hline
Workers Intensity ($\nu_w$)   & $7\times 10^{-4}$ $\text{m}^{-2}$ \\
\hline
Requester Intensity ($\nu_r$)  & $1\times 10^{-4}$ $\text{m}^{-2}$\\
\hline
Carrier frequency & $28~\mathrm{GHz}$  \\
\hline
LoS and NLoS path loss exponent ($\alpha_L, \alpha_N$) & 2, 4 \cite{yu2017coverage} \\
\hline
Fading values for LoS and NLoS ($N_L, N_N$) & 3, 2 \cite{yu2017coverage} \\
\hline
Path loss intercepts ($C_L$, $C_N$) & $-61.4 \, \text{dB}$, $-72 \, \text{dB}$ \cite{yu2017coverage} \\ 
\hline
Main lobe gains ($M_{w}$ = $M_{r}$) & 5 dBi \cite{chen2019coverage} \\
\hline
Side lobe gains ($m_{w}$ = $m_{r}$) & -5 dBi \cite{chen2019coverage}\\
\hline
SINR threshold ($\xi$) & 5 dB \cite{yu2017coverage} \\
\hline
3-dB beamwidth ($\theta_{r}=\theta_{w}$)  & 45\degree \cite{andrews2016modeling}\\
\hline
Bandwidth ($B$) & $200~\mathrm{MHz}$ \cite{andrews2016modeling} \\
\hline
Transmit power ($P_t$) & $30~\mathrm{dBm}$ \cite{andrews2016modeling} \\
\hline
Noise figure & $10~\mathrm{dB}$ \cite{andrews2016modeling} \\
\hline
\parbox[t]{0.62\linewidth}{Thermal noise power $\left(\sigma_{\mathrm{th}}^{2}= -174~\mathrm{dBm/Hz}\right.$\\
$\left.+10\log_{10}\!\big(B[\mathrm{Hz}]\big)\right.$\\
$\left.+10~\mathrm{dB}~(\text{noise figure})\right)$\cite{chen2019coverage}}
& $-81~\mathrm{dBm}$ \\
\hline
Normalized noise $\sigma^{2}= \sigma_{\mathrm{th}}^{2}/P_t$ &
$-111~\mathrm{dB}$ \\
\hline
Task execution rate ($\mu_f$) & 0.02 task/second \\
\hline
D2D communication time ($\tau_c$) & 1 second \\
\hline
Maximum radius for LoS devices ($R_L$) & 100 m \\
\hline
Reliability parameter ($l$)  & 3\\
\hline
\end{tabular}}
\end{table}

This section presents numerical and simulation results to validate the proposed spatiotemporal models and evaluate their performance across a wide range of operating conditions. We first confirm the accuracy of the analytical framework and then conduct  sensitivity analysis that systematically explores the impact of key system parameters to assess the robustness of the observed trends. Unless otherwise stated, the default network parameters are listed in Table~\ref{table_params}. The key mmWave communication parameter values are selected in line with well-established mmWave studies in the literature.  We emphasize that the developed mathematical model remains valid across a broad range of parameter values; the specific choices are adopted to demonstrate a typical operating conditions.

The Monte Carlo simulations are conducted over an area of $10~\text{km}^2$, where requesters and EED workers are generated as independent PPPs, and a typical requester is fixed at the origin. Around this requester, a disk of radius $R_L$ defines the region of potential LoS links; devices inside this disk are tagged as LoS and those outside as NLoS. For every D2D link, we apply the distance-dependent path-loss model and sample channel fading gains from the corresponding Gamma distribution. Directional beamforming gain is computed from the actual link angles at the transmitter and receiver by selecting one of the four possible combined gains specified in Table~\ref{directivity_gain}. The typical requester offloads to LoS workers within radius $R_L$ according to the considered policy (random or location-aware); for each candidate worker, we compute its received SINR, including aggregate interference from LoS and NLoS requesters, and declare the offloading attempt successful if the SINR exceeds the threshold $\xi$. Repeating this over $10^5$ independent network realizations yields the successful offloading probabilities of the first, second, and subsequent ordered workers, which are then converted into offloading rates.

Given a segmentation level $n$, we next simulate the ACTMC-based queueing model to obtain the end-to-end task response delay. The task is split into $n$ equal segments, and we simultaneously track (i) sequential offloading, where an additional segment is assigned to a worker with an offloading rate determined following the procedure described previously, and (ii) parallel execution of the offloaded segments, which complete with exponential service times of mean $1/(n\mu_f)$. Starting from an initial configuration where no segment has yet been offloaded, we track the time until all $n$ segments have completed, which gives one realization of the end-to-end task response delay. Averaging these delays over $10^5$ realizations yields the simulated average task response delay.

\begin{figure}[h]
\centering
\begin{tikzpicture}[scale=0.7, transform shape,font=\Large]

\definecolor{color0}{rgb}{0.12156862745098,0.466666666666667,0.705882352941177}
\definecolor{mycolor1}{rgb}{0.00000,0.44700,0.74100}%
\definecolor{mycolor2}{rgb}{0.85000,0.32500,0.09800}%
\definecolor{mycolor3}{rgb}{0.92900,0.69400,0.12500}%
\definecolor{mycolor4}{rgb}{0.49400,0.18400,0.55600}%
\definecolor{mycolor5}{rgb}{0.46600,0.67400,0.18800}%
\definecolor{mycolor6}{rgb}{0.30100,0.74500,0.93300}%
\definecolor{mycolor7}{rgb}{0.63500,0.07800,0.18400}%

\begin{axis}[
width=3.8in,
height=3.2in,
legend cell align={left},
legend style={at={(0.465,0.35)},font=\small, fill opacity=0.8, draw opacity=1, text opacity=1, draw=white!80!black},
tick align=outside,
tick pos=left,
x grid style={white!69.0196078431373!black},
xlabel={SINR Threshold $\xi$ (dB)},
xmin=-15, xmax=10,
xtick={-20,-15,-10,...,15},
xtick style={color=black},
y grid style={white!69.0196078431373!black},
ylabel={Offloading Success Probability},
ymin=0, ymax=1,
ytick style={color=black},
xmajorgrids,
ymajorgrids
]

\addplot [line width=0.6mm, mycolor1]
  table[row sep=crcr]{%
-20	0.998978084751832\\
-19	0.998709032509371\\
-18	0.998368607184697\\
-17	0.99793757111153\\
-16	0.997391417776439\\
-15	0.996698937067623\\
-14	0.995820426555203\\
-13	0.994705498281431\\
-12	0.993290452141604\\
-11	0.99149523157173\\
-10	0.98922005184349\\
-9	0.986341896650486\\
-8	0.982711201820965\\
-7	0.9781491514976\\
-6	0.972446047657215\\
-5	0.96536112838692\\
-4	0.956624013478574\\
-3	0.945937783249096\\
-2	0.932983817417596\\
-1	0.917429193116857\\
0	0.898938600215922\\
1	0.877193711744101\\
2	0.851922591498399\\
3	0.822939082549709\\
4	0.790187407627961\\
5	0.753782215719643\\
6	0.714031755917012\\
7	0.671433939875595\\
8	0.626642008924347\\
9	0.580406128401667\\
10	0.533505494004363\\
11	0.486688401995572\\
12	0.440633300818381\\
13	0.395933924512048\\
14	0.353101291257539\\
15	0.312570054336428\\
};
\addlegendentry{$R_L$ = 100 (random)}

\addplot [line width=0.6mm, mycolor2]
  table[row sep=crcr]{%
-20	0.989649247975225\\
-19	0.986633972245691\\
-18	0.982686370813717\\
-17	0.977505708993583\\
-16	0.970699390502465\\
-15	0.961763872573421\\
-14	0.950069335151775\\
-13	0.9348547843723\\
-12	0.915243097415979\\
-11	0.890287349839792\\
-10	0.859058726806078\\
-9	0.820780208000808\\
-8	0.774997654976903\\
-7	0.72176209784879\\
-6	0.661779108099074\\
-5	0.596472011923537\\
-4	0.527914490195832\\
-3	0.458618502488185\\
-2	0.391208711540644\\
-1	0.328057532134825\\
0	0.270975045735328\\
1	0.221032661007684\\
2	0.1785528308943\\
3	0.143240435254956\\
4	0.114391193448824\\
5	0.091104750738164\\
6	0.0724514829364216\\
7	0.057574837460459\\
8	0.045737041566027\\
9	0.036327318387592\\
10	0.0288509864993871\\
11	0.0229118146680068\\
12	0.0181940384679722\\
13	0.0144465472137395\\
14	0.0114698022255968\\
15	0.00910528697836459\\
};
\addlegendentry{$R_L$ = 300 (random)}

\addplot [line width=0.6mm, mycolor4]
  table[row sep=crcr]{%
-20	0.99965057450611\\
-19	0.99962662110339\\
-18	0.999596439066847\\
-17	0.999558398216205\\
-16	0.999510436506497\\
-15	0.999449944420754\\
-14	0.999373617442141\\
-13	0.999277267621197\\
-12	0.999155582789248\\
-11	0.999001818898824\\
-10	0.998807407264489\\
-9	0.998561454200156\\
-8	0.998250105950443\\
-7	0.997855747446076\\
-6	0.997356000281588\\
-5	0.996722485021865\\
-4	0.995919317738477\\
-3	0.99490132323366\\
-2	0.993611970198083\\
-1	0.991981067692025\\
0	0.989922306234086\\
1	0.987330775250424\\
2	0.984080633926391\\
3	0.9800231482917\\
4	0.974985335279428\\
5	0.968769489402098\\
6	0.961153935049935\\
7	0.951895466862408\\
8	0.940734097789803\\
9	0.927400855908396\\
10	0.911629325801791\\
11	0.893171275374018\\
12	0.871815972185783\\
13	0.847411756615805\\
14	0.819887368155209\\
15	0.789269804059909\\
};
\addlegendentry{k=1, $R_L=100$m}

\addplot [line width=0.6mm, mycolor5]
  table[row sep=crcr]{%
-20	0.997587504759054\\
-19	0.997564355937188\\
-18	0.997534814718275\\
-17	0.997497198272993\\
-16	0.99744940006583\\
-15	0.997388770047437\\
-14	0.997311953925739\\
-13	0.997214680850841\\
-12	0.997091485899306\\
-11	0.996935348228305\\
-10	0.996737217082059\\
-9	0.996485385695993\\
-8	0.996164657225211\\
-7	0.995755226306696\\
-6	0.99523117351391\\
-5	0.994558437022758\\
-4	0.99369208748215\\
-3	0.992572693186185\\
-2	0.99112153353265\\
-1	0.98923441823587\\
0	0.986773929101853\\
1	0.983560066785263\\
2	0.979359614988757\\
3	0.973875084289866\\
4	0.966734890174586\\
5	0.957487401613199\\
6	0.945602486996083\\
7	0.930484839951336\\
8	0.911503201719871\\
9	0.888038097742377\\
10	0.859547568300076\\
11	0.825645784151633\\
12	0.78618428574237\\
13	0.741321409756745\\
14	0.691564056011227\\
15	0.637768626293979\\
};
\addlegendentry{k=1, $R_L=300$m} 

\addplot [
  semithick, black, mark=x, thick,
  mark size=4pt, mark options={solid},
  only marks,
  each nth point=3
]
  table[row sep=crcr]{%
-20	0.9988\\
-19	0.99835\\
-18	0.99795\\
-17	0.99755\\
-16	0.997\\
-15	0.99665\\
-14	0.99595\\
-13	0.99495\\
-12	0.9937\\
-11	0.99215\\
-10	0.9898\\
-9	0.9866\\
-8	0.98305\\
-7	0.97845\\
-6	0.9728\\
-5	0.96535\\
-4	0.9567\\
-3	0.94535\\
-2	0.9307\\
-1	0.91495\\
0	0.89625\\
1	0.87285\\
2	0.84745\\
3	0.81835\\
4	0.78625\\
5	0.74775\\
6	0.70505\\
7	0.66285\\
8	0.617\\
9	0.5726\\
10	0.52525\\
11	0.47705\\
12	0.42955\\
13	0.38525\\
14	0.3425\\
15	0.30155\\
};
\addlegendentry{Simulations}
\addplot [
  semithick, black, mark=x, thick,
  mark size=4pt, mark options={solid},
  only marks,
  each nth point=3
]
  table[row sep=crcr]{%
-20	0.98985\\
-19	0.9861\\
-18	0.98185\\
-17	0.977\\
-16	0.9708\\
-15	0.96085\\
-14	0.94925\\
-13	0.93325\\
-12	0.9124\\
-11	0.8864\\
-10	0.8519\\
-9	0.8117\\
-8	0.7615\\
-7	0.705\\
-6	0.64295\\
-5	0.576\\
-4	0.50595\\
-3	0.4345\\
-2	0.3689\\
-1	0.30905\\
0	0.2505\\
1	0.20205\\
2	0.16175\\
3	0.12755\\
4	0.10225\\
5	0.0821\\
6	0.0644\\
7	0.05055\\
8	0.04005\\
9	0.03205\\
10	0.0258\\
11	0.0194\\
12	0.0157\\
13	0.0126\\
14	0.0106\\
15	0.00925\\
};

\addplot [
  semithick, black, mark=x, thick,
  mark size=4pt, mark options={solid},
  only marks,
  each nth point=3
]
  table[row sep=crcr]{%
-20	1\\
-19	0.99995\\
-18	0.99985\\
-17	0.99985\\
-16	0.99975\\
-15	0.9997\\
-14	0.9996\\
-13	0.99955\\
-12	0.9995\\
-11	0.9993\\
-10	0.99895\\
-9	0.9985\\
-8	0.9984\\
-7	0.99795\\
-6	0.9973\\
-5	0.9966\\
-4	0.9957\\
-3	0.9945\\
-2	0.9927\\
-1	0.9908\\
0	0.98865\\
1	0.9861\\
2	0.9824\\
3	0.97885\\
4	0.97385\\
5	0.96735\\
6	0.9604\\
7	0.94985\\
8	0.93795\\
9	0.92345\\
10	0.90475\\
11	0.88615\\
12	0.86355\\
13	0.8388\\
14	0.8094\\
15	0.7766\\
};

\addplot [
  semithick, black, mark=x, thick,
  mark size=4pt, mark options={solid},
  only marks,
  each nth point=3
]
  table[row sep=crcr]{%
-20	0.9999\\
-19	0.9999\\
-18	0.99985\\
-17	0.9998\\
-16	0.99975\\
-15	0.9997\\
-14	0.9997\\
-13	0.99955\\
-12	0.99955\\
-11	0.99935\\
-10	0.9992\\
-9	0.9991\\
-8	0.9985\\
-7	0.99795\\
-6	0.9973\\
-5	0.99645\\
-4	0.9958\\
-3	0.9944\\
-2	0.9931\\
-1	0.99125\\
0	0.9891\\
1	0.986\\
2	0.982\\
3	0.9761\\
4	0.9675\\
5	0.9595\\
6	0.94585\\
7	0.92945\\
8	0.9092\\
9	0.88525\\
10	0.85745\\
11	0.8201\\
12	0.77925\\
13	0.7332\\
14	0.6809\\
15	0.626\\
};
\end{axis}

\end{tikzpicture} 
 \vspace{-15pt} 
\caption{Average D2D offloading success probability vs SINR threshold $\xi$.}
\label{fig:SINR_validation}
\end{figure}
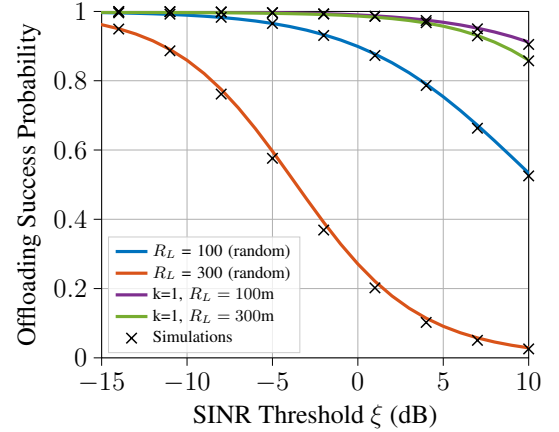
\setlength{\textfloatsep}{10pt}

Fig.~\ref{fig:SINR_validation} shows the average successful offloading probability $p_s$ for the random and ordered EED-selection scenarios as a function of the SINR threshold $\xi$ for different values of the LoS radius $R_L$. The case $k=1$ corresponds to offloading to the nearest EED relative to the requester. The close match between the Monte Carlo simulations and the proposed analytical expressions confirms the accuracy of Theorems~1 and~3 over a wide range of $\xi$ and $R_L$ values.
As expected, $p_s$ decreases monotonically with $\xi$ because a higher SINR threshold imposes a stricter link-quality requirement, making successful offloading less likely. The figure also reveals a strong sensitivity to the EED-selection strategy: offloading to the nearest device ($k=1$) consistently achieves a much higher success probability than random selection, owing to the shorter average distance between the requester and the worker and, hence, a stronger received signal power. Furthermore, varying $R_L$ illustrates the impact of the propagation environment. Larger $R_L$ values increase the likelihood of longer LoS D2D links and simultaneously expose the requester to stronger interference from other LoS transmitters, both of which reduce the successful offloading probability.

\begin{figure}[t]
\centering
\begin{tikzpicture}[scale=0.7, transform shape, font=\Large]

\definecolor{color0}{rgb}{0.12156862745098,0.466666666666667,0.705882352941177}
\definecolor{mycolor1}{rgb}{0.00000,0.44700,0.74100}%
\definecolor{mycolor2}{rgb}{0.85000,0.32500,0.09800}%
\definecolor{mycolor3}{rgb}{0.92900,0.69400,0.12500}%
\definecolor{mycolor4}{rgb}{0.49400,0.18400,0.55600}%
\definecolor{mycolor5}{rgb}{0.46600,0.67400,0.18800}%
\definecolor{mycolor6}{rgb}{0.30100,0.74500,0.93300}%
\definecolor{mycolor7}{rgb}{0.63500,0.07800,0.18400}%

\begin{axis}[
width=3.8in,
height=3.2in,
legend cell align={left},
legend style={
  at={(0.28,0.79)},
  anchor=north west,
  fill opacity=0.8,
  draw opacity=1,
  text opacity=1,
  draw=white!80!black,
  font=\small,
  inner sep=1pt,
  row sep=1pt,
  column sep=2pt,
  cells={anchor=west}
},
tick align=outside,
tick pos=left,
x grid style={white!69.0196078431373!black},
xlabel={Number of Task Segments ($n$)},
xmin=1, xmax=12,
xtick={0,1,2,...,12},
xtick style={color=black},
y grid style={white!69.0196078431373!black},
ylabel={Average Response Delay (sec)},
ymin=25, ymax=45,
  ytick={5,10,15,20,25,30,35,40,45,50,55,60},
ytick style={color=black},
xmajorgrids,
ymajorgrids
]

\addplot [line width=0.6mm, color=mycolor1]
table {
1 54.045218151715
2 41.2691949724367
3 36.7437943523897
4 33.3997768087473
5 32.0444135512486
6 31.6992450668801
7 32.3432660309726
8 33.0070827116297
9 33.9495292531741
10 35.1407933209938
11 36.8041991135726
12 38.4776706120547
};
\addlegendentry{Random EEDs}

\addplot [line width=0.6mm, mycolor2]
table {
1 51.0089901018071
2 40.0164420302808
3 34.6140312550207
4 31.671075453434
5 30.0636428750901
6 29.2788689902954
7 29.045840479324
8 29.2060908855222
9 29.6593446547928
10 30.3378312327232
11 31.193034592938
12 32.1884550138858
};
\addlegendentry{Ordered EEDs}

\addplot [color=black, draw=none, mark=x, thick, mark size=4pt, mark options={black}, only marks]
table {
1 52.6537869539833
2 41.5682827742797
3 36.194278628613
4 33.3405711902318
5 31.9619097964449
6 31.5155401681403
7 31.9699199322219
8 32.7226226588997
9 33.6311994290021
10 35.0521118239123
11 36.8007372041963
12 38.6390101337421
};
\addlegendentry{Simulations}

\addplot [color=black, draw=none, mark=x, thick, mark size=4pt, mark options={black}, only marks]
table {
1 50.4385381728588
2 40.2673227185931
3 34.6667169239198
4 31.6511534965235
5 30.242823907063
6 29.2553854540296
7 29.0749583522179
8 29.2482942453521
9 29.6891094037183
10 30.3376065408027
11 31.196025548563
12 32.0394017657581
};

\end{axis}
\end{tikzpicture}
 \vspace{-15pt} 
\caption{Average task response delay vs the number of task segments.}
\label{fig:time_analysis}
\end{figure}
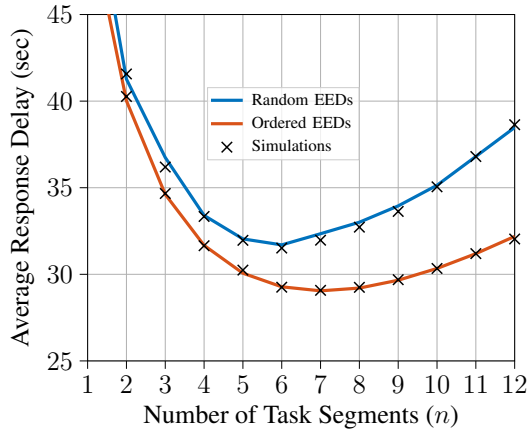

Fig.~\ref{fig:time_analysis} depicts the average task response delay as a function of the number of allocated workers (segments) $n$. The close agreement between the Monte Carlo simulations and the proposed analytical expressions validates Theorem~2 over a wide range of $n$ values. Consistent with Fig.~\ref{fig:SINR_validation}, selecting EEDs based on their distance from the requester significantly reduces the average response delay compared to random selection. This improvement stems from reduced offloading time, which results from the higher probability of successful offloading when associating with closer devices.

The figure also illustrates the sensitivity of the average delay to the segmentation level $n$, revealing a key design insight: an optimal number of segments $n$ exists that minimizes the average task response delay. This behavior directly manifests the fundamental trade-off between communication and computation. For small to moderate $n$, increasing the number of segments reduces computation time because more workers share the task, while the additional communication overhead remains limited, thereby decreasing the overall delay. This trend continues until the communication delay, driven by longer links and more frequent retransmissions, becomes dominant and outweighs the reduction in computation delay, causing the average task response delay to increase.

\setlength{\textfloatsep}{10pt}

\subsection{Baseline Model Results}
After validating the theoretical analysis against simulations, we conduct a sensitivity analysis to examine how key system parameters affect the performance of the baseline model.

\begin{figure*}[t]
    \centering
    \subfloat[Over varying task execution rate ($\mu_{f}$) \label{fig:subfig-a}]
    {

\begin{tikzpicture}[scale=0.81, transform shape,font=\Large]
\definecolor{color0}{rgb}{0.12156862745098,0.466666666666667,0.705882352941177}

\definecolor{color1}{rgb}{0.00000,0.44700,0.74100}%
\definecolor{color2}{rgb}{0.85000,0.32500,0.09800}%
\definecolor{color3}{rgb}{0.92900,0.69400,0.12500}%
\definecolor{color4}{rgb}{0.49400,0.18400,0.55600}%
\definecolor{color5}{rgb}{0.46600,0.67400,0.18800}%
\definecolor{color6}{rgb}{0.30100,0.74500,0.93300}%
\definecolor{color7}{rgb}{0.63500,0.07800,0.18400}%

\begin{axis}[
width=2.6in,
height=2.6in,
legend cell align={left},
legend style={font=\small,fill opacity=0.8, draw opacity=1, text opacity=1, draw=white!80!black},
tick align=outside,
tick pos=left,
x grid style={white!69.0196078431373!black},
xlabel={\normalsize{Number of allocated  Workers (n)}},
xmin=1, xmax=10.1,
xtick style={color=black},
y grid style={white!69.0196078431373!black},
ylabel={\normalsize{Average Response Delay (sec)}},
ymin=0, ymax=200,
ytick style={color=black},
xtick={0,1,2,...,10},
xmajorgrids,
ymajorgrids
]
\addplot [line width=0.6mm, color1]
table {%
1 202.653786953983
2 153.818582300879
3 127.653937878476
4 110.899580298552
5 99.5530892677754
6 91.483014925214
7 85.4084705082734
8 80.9072688669775
9 77.584646279476
10 75.0709588465919
};
\addlegendentry{$\mu_f$ = 0.005}
\addplot [dashed, line width=0.9mm, black]
table {%
1 102.653786953983
2 79.0565105741486
3 66.6090647127553
4 59.0191834583541
5 54.2414158224936
6 51.0986460850656
7 49.0844583270957
8 47.9469179354952
9 47.398521804365
10 47.4145162279978
};
\addlegendentry{$\mu_f$ = 0.01}
\addplot [dash dot, line width=0.9mm, color4]
table {%
1 22.6537869539833
2 19.2583096796107
3 18.274107020434
4 18.5038855297196
5 19.4729800208616
6 20.8761431327055
7 22.711306556275
8 24.6770904052429
9 26.8100808252077
10 29.2377015501171
};
\addlegendentry{$\mu_f$ = 0.05}
\addplot [dotted, line width=0.9mm, color5]
table {%
1 13.7648980650944
2 12.7433557268453
3 13.2312589554942
4 14.524464152741
5 16.2350605867722
6 18.3531188796346
7 20.6496799957369
8 22.9949150151513
9 25.3347296581914
10 27.7838264424802
};
\addlegendentry{$\mu_f$ = 0.1}
\addplot [semithick, red, mark=*, mark size=2.0, mark options={solid}, only marks, forget plot]
table {%
10 75.0709588465919
};
\addplot [semithick, red, mark=*, mark size=2.0, mark options={solid}, only marks, forget plot]
table {%
9 47.398521804365
};
\addplot [semithick, red, mark=*, mark size=2.0, mark options={solid}, only marks, forget plot]
table {%
3 18.274107020434
};
\addplot [semithick, red, mark=*, mark size=2.0, mark options={solid}, only marks, forget plot]
table {%
2 12.7433557268453
};
\end{axis}
\end{tikzpicture}


}%
    \hspace{0.035\textwidth}
    \subfloat[\shortstack{Over varying execution and offloading rates ratio} \label{fig:subfig-b}]
    {
\begin{tikzpicture}[scale=0.81, transform shape,font=\Large]

\definecolor{color0}{rgb}{0.12156862745098,0.466666666666667,0.705882352941177}

\definecolor{color1}{rgb}{0.00000,0.44700,0.74100}%
\definecolor{color2}{rgb}{0.85000,0.32500,0.09800}%
\definecolor{color3}{rgb}{0.92900,0.69400,0.12500}%
\definecolor{color4}{rgb}{0.49400,0.18400,0.55600}%
\definecolor{color5}{rgb}{0.46600,0.67400,0.18800}%
\definecolor{color6}{rgb}{0.30100,0.74500,0.93300}%
\definecolor{color7}{rgb}{0.63500,0.07800,0.18400}%

\begin{axis}[
width=2.6in,
height=2.6in,
legend cell align={left},
legend style={font=\small,at={(0.85,0.98)}, fill opacity=0.8, draw opacity=1, text opacity=1, draw=white!80!black},
tick align=outside,
tick pos=left,
x grid style={white!69.0196078431373!black},
xlabel={\normalsize{Number of allocated  Workers (n)}},
xmin=1, xmax=10,
xtick={1,2,...,10},
xtick style={color=black},
y grid style={white!69.0196078431373!black},
ymin=0, ymax=100,
ytick style={color=black},
xmajorgrids,
ymajorgrids
]
\addplot [line width=0.6mm, color1]
table {%
1 101.967226014597
2 77.9386714763568
3 65.0517983925334
4 57.1874198085011
5 51.9179544276406
6 48.2598894334428
7 45.6963663641472
8 43.9542293703741
9 42.8033423767706
10 42.1071070210023
};
\addlegendentry{$\nicefrac{\mu_f}{\lambda_h}  = 0.01 $}
\addplot [dashed, line width=0.9mm, black]
table {%
1 21.9672260145968
2 18.1258268007385
3 16.5896108295952
4 16.1861921388522
5 16.4729798011619
6 17.1894820240264
7 18.2634084636453
8 19.5198686740339
9 20.9987384831627
10 22.5800053400806
};
\addlegendentry{$\nicefrac{\mu_f}{\lambda_h}  = 0.05 $}
\addplot [dash dot, line width=0.9mm, color4]
table {%
1 11.9672260145968
2 10.7298206481962
3 10.7802216490894
4 11.5129355021567
5 12.6835184362655
6 14.0632002383478
7 15.6571829812226
8 17.3390681952943
9 19.115764762475
10 20.8801888358258
};
\addlegendentry{$\nicefrac{\mu_f}{\lambda_h}  = 0.1$}
\addplot [dotted, line width=0.9mm, color5]
table {%
1 2.96722601459682
2 4.48967845241597
3 6.27981317613524
4 8.15014232446527
5 10.0629739604116
6 11.9589169073432
7 13.8882856995727
8 15.863829958511
9 17.8716522136886
10 19.7498658932702
};

\addlegendentry{$\nicefrac{\mu_f}{\lambda_h}  = 1$}
\addplot [semithick, red, mark=*, mark size=2, mark options={solid}, only marks, forget plot]
table {%
1 2.96722601459682
};
\addplot [semithick, red, mark=*, mark size=2, mark options={solid}, only marks, forget plot]
table {%
2 10.7298206481962
};
\addplot [semithick, red, mark=*, mark size=2, mark options={solid}, only marks, forget plot]
table {%
4 16.1861921388522
};
\addplot [semithick, red, mark=*, mark size=2, mark options={solid}, only marks, forget plot]
table {%
10 42.1071070210023
};
\end{axis}

\end{tikzpicture}
}%
    \hspace{0.035\textwidth}
    \subfloat[Over varying SINR threshold ($\xi$) \label{fig:subfig-c}]
    {
\begin{tikzpicture}[scale=0.81, transform shape,font=\Large]

\definecolor{color0}{rgb}{0.12156862745098,0.466666666666667,0.705882352941177}

\definecolor{color1}{rgb}{0.00000,0.44700,0.74100}%
\definecolor{color2}{rgb}{0.85000,0.32500,0.09800}%
\definecolor{color3}{rgb}{0.92900,0.69400,0.12500}%
\definecolor{color4}{rgb}{0.49400,0.18400,0.55600}%
\definecolor{color5}{rgb}{0.46600,0.67400,0.18800}%
\definecolor{color6}{rgb}{0.30100,0.74500,0.93300}%
\definecolor{color7}{rgb}{0.63500,0.07800,0.18400}%

\begin{axis}[
width=2.6in,
height=2.6in,
legend cell align={left},
legend style={font=\small,at={(0.55,0.98)}, fill opacity=0.8, draw opacity=1, text opacity=1, draw=white!80!black},
tick align=outside,
tick pos=left,
x grid style={white!69.0196078431373!black},
xlabel={\normalsize{Number of allocated  Workers (n)}},
xmin=1, xmax=10.1,
xtick style={color=black},
y grid style={white!69.0196078431373!black},
ymin=0, ymax=300,
ytick={0,50,100,...,300},
ytick style={color=black},
xtick={0,1,2,...,10},
xmajorgrids,
ymajorgrids
]
\addplot [line width=0.6mm, color1]
table {%
1 75.5362614913177
2 82.3173976007528
3 97.7622178807476
4 117.264527126927
5 139.706348373178
6 162.371945122563
7 186.70755863602
8 210.652217020417
9 235.634531980709
10 261.647741274086
};
\addlegendentry{$\xi$ = 0}
\addplot [dashed, line width=0.9mm, black]
table {%
1 58.0749354005168
2 50.6164355216408
3 49.3499104967646
4 51.2285225764658
5 55.2638927001246
6 60.3000901451397
7 66.3134726400855
8 72.7806277192635
9 79.6994803407215
10 87.220156509418
};
\addlegendentry{$\xi$ = -5}
\addplot [dash dot, line width=0.9mm, color4]
table {%
1 52.6537869539833
2 41.5682827742797
3 36.194278628613
4 33.3405711902318
5 31.9619097964449
6 31.5155401681403
7 31.7699199322219
8 32.5226226588997
9 33.6311994290021
10 35.0521118239123
};
\addlegendentry{$\xi$ = -10}
\addplot [dotted, line width=0.9mm, color5]
table {%
1 51.0531193394836
2 39.105794256561
3 32.7239583518548
4 28.7963057552298
5 26.1841739531327
6 24.4138023670665
7 23.1883221212071
8 22.3668681212094
9 21.8537657271933
10 21.5850934590613
};
\addlegendentry{$\xi$ = -20}

\addplot [semithick, red, mark=*, mark size=2, mark options={solid}, only marks, forget plot]
table {%
10 21.5850934590613
};
\addplot [semithick, red, mark=*, mark size=2, mark options={solid}, only marks, forget plot]
table {%
6 31.5155401681403
};
\addplot [semithick, red, mark=*, mark size=2, mark options={solid}, only marks, forget plot]
table {%
3 49.3499104967646
};
\addplot [semithick, red, mark=*, mark size=2, mark options={solid}, only marks, forget plot]
table {%
1 75.5362614913177
};
\end{axis}

\end{tikzpicture}}
    \caption{Average task response delay ($T_A$) vs the number of allocated workers ($n$) for different system parameters.}
    \label{fig:params}
\end{figure*}
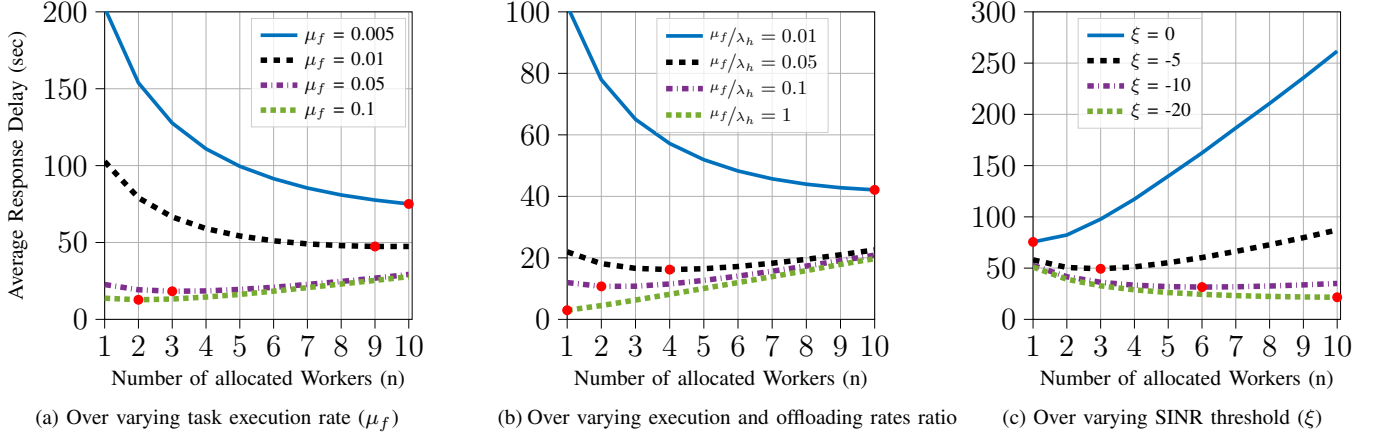

Fig.~\ref{fig:params} provides a sensitivity analysis of the optimal segmentation level with respect to key system parameters. In all subfigures, the red dots indicate the minimum average task response delay, i.e., the operating point where the system allocates the optimal number of workers $n$.
In Fig.~\ref{fig:params}(a), we observe that the optimal number of workers strongly depends on the task execution rate $\mu_f$. When $\mu_f$ is small, each worker is relatively slow, so involving more workers is beneficial to reduce the computation time, which shifts the optimal point to larger $n$. As $\mu_f$ increases, each worker can process tasks faster, and fewer workers are needed to minimize the response delay. Beyond the optimal $n$, adding more workers only increases the offloading overhead and the likelihood of retransmissions, which leads to performance degradation and explains the rise of the curves after the red dots.

Fig.~\ref{fig:params}(b) explores the impact of the ratio $\nicefrac{\mu_f}{\lambda_h}$, where $\lambda_h$ is the average offloading rate. When $\mu_f$ is low relative to $\lambda_h$, the system is computation-limited: computation dominates the total delay, and more workers are required to reduce the response time. As $\mu_f$ increases relative to $\lambda_h$, the benefit of parallelism diminishes and the optimal $n$ decreases. The case $\nicefrac{\mu_f}{\lambda_h}=1$ corresponds to a regime where communication and computation delays are of the same order; in this case, offloading to a single worker is sufficient, as further segmentation would primarily increase communication costs without providing meaningful computation gains.
Finally, Fig.~\ref{fig:params}(c) shows the effect of the SINR threshold $\xi$. Increasing $\xi$ tightens the link-quality requirement, which reduces the successful offloading probability and increases the expected communication delay. In this communication-limited regime, it becomes preferable to use fewer workers to avoid excessive offloading overhead, and thus the optimal $n$ shifts to smaller values. Overall, this figure highlights how the optimal segmentation level is highly sensitive to these main system parameters and provides practical guidelines for tuning $n$ under different operating conditions.

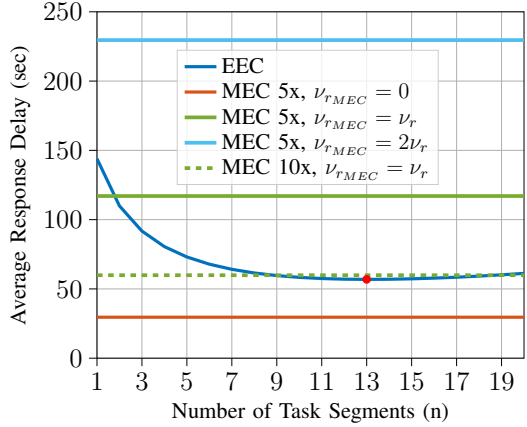
\begin{figure}[t]
\centering

\begin{tikzpicture}[scale=0.7, transform shape,font=\Large]

\definecolor{color0}{rgb}{0.12156862745098,0.466666666666667,0.705882352941177}
\definecolor{color1}{rgb}{0.00000,0.44700,0.74100}%
\definecolor{color2}{rgb}{0.85000,0.32500,0.09800}%
\definecolor{color3}{rgb}{0.92900,0.69400,0.12500}%
\definecolor{color4}{rgb}{0.49400,0.18400,0.55600}%
\definecolor{color5}{rgb}{0.46600,0.67400,0.18800}%
\definecolor{color6}{rgb}{0.30100,0.74500,0.93300}%
\definecolor{color7}{rgb}{0.63500,0.07800,0.18400}%

\begin{axis}[
width=3.8in,
height=3.2in,
legend cell align={left},
legend style={at={(0.8,0.89)}, font=\large, fill opacity=0.8, draw opacity=1, text opacity=1, draw=white!80!black},
grid=both,
tick align=outside,
tick pos=left,
x grid style={white!69.0196078431373!black},
xlabel={\large{Number of Task Segments (n)}},
xmin=1, xmax=20,
xtick style={color=black},
y grid style={white!69.0196078431373!black},
ylabel={\large{Average Response Delay (sec)}},
ymin=0, ymax=250,
ytick style={color=black},
xtick={1, 3, ..., 20},
xmajorgrids,
ymajorgrids
]

\addplot [line width=0.6mm, color1]
table {%
1 143.968537049026
2 109.912828715512
3 91.74172130162
4 80.5237665558168
5 73.046446594841
6 67.8435372922055
7 64.142037153168
8 61.4926213626342
9 59.6149497505365
10 58.3244309215533
11 57.494225723462
12 57.0340797766108
13 56.8778385471824
14 56.9757307054826
15 57.2894099900306
16 57.7886656321705
17 58.4491823125262
18 59.2509836429946
19 60.1773346829175
20 61.2139608574658
};
\addlegendentry{EEC}

\addplot [line width=0.6mm, color2]
table {%
1 29.5720048806247
2 29.5720048806247
3 29.5720048806247
4 29.5720048806247
5 29.5720048806247
6 29.5720048806247
7 29.5720048806247
8 29.5720048806247
9 29.5720048806247
10 29.5720048806247
11 29.5720048806247
12 29.5720048806247
13 29.5720048806247
14 29.5720048806247
15 29.5720048806247
16 29.5720048806247
17 29.5720048806247
18 29.5720048806247
19 29.5720048806247
20 29.5720048806247
};
\addlegendentry{MEC 5x, $\nu_{r_{MEC}} = 0$}

\addplot [line width=0.7mm, color5]
table {%
1 117.038046516936
2 117.038046516936
3 117.038046516936
4 117.038046516936
5 117.038046516936
6 117.038046516936
7 117.038046516936
8 117.038046516936
9 117.038046516936
10 117.038046516936
11 117.038046516936
12 117.038046516936
13 117.038046516936
14 117.038046516936
15 117.038046516936
16 117.038046516936
17 117.038046516936
18 117.038046516936
19 117.038046516936
20 117.038046516936
};
\addlegendentry{MEC 5x, $\nu_{r_{MEC}} = \nu_r$}

\addplot [line width=0.7mm, color6]
table {%
1 229.572004880625
2 229.572004880625
3 229.572004880625
4 229.572004880625
5 229.572004880625
6 229.572004880625
7 229.572004880625
8 229.572004880625
9 229.572004880625
10 229.572004880625
11 229.572004880625
12 229.572004880625
13 229.572004880625
14 229.572004880625
15 229.572004880625
16 229.572004880625
17 229.572004880625
18 229.572004880625
19 229.572004880625
20 229.572004880625
};
\addlegendentry{MEC 5x, $\nu_{r_{MEC}} = 2\nu_r$}


\addplot [dashed, line width=0.7mm, color5]
table {%
1 59.8951893740789
2 59.8951893740789
3 59.8951893740789
4 59.8951893740789
5 59.8951893740789
6 59.8951893740789
7 59.8951893740789
8 59.8951893740789
9 59.8951893740789
10 59.8951893740789
11 59.8951893740789
12 59.8951893740789
13 59.8951893740789
14 59.8951893740789
15 59.8951893740789
16 59.8951893740789
17 59.8951893740789
18 59.8951893740789
19 59.8951893740789
20 59.8951893740789
};
\addlegendentry{MEC 10x, $\nu_{r_{MEC}} = \nu_r$}


\addplot [semithick, red, mark=*, mark size=2.0, mark options={solid}, only marks, forget plot]
table {%
13 56.8778385471824
};

\end{axis}
\end{tikzpicture}

\caption{MEC and EEC average response delay using varying MEC congestion scenarios.}
\label{results_MEC_and_EEDs_time}
\end{figure}

To test the performance of our proposed EEC framework, we conduct a comparison against centralized MEC systems under various computational and congestion scenarios. The MEC employs an advanced parallel architecture \cite{lit13,ndikumana2019joint} where a single physical machine (PM) hosts multiple VMs for concurrent task processing. In all MEC scenarios, requesters within the LoS radius $R_L$ offload complete tasks without partitioning, with congestion defined by the number of concurrent requesters served.
We analyze a practical MEC that possesses computational power five times superior to that of an EED, which is consistent with typical values reported in literature \cite{ndikumana2019joint}. We also analyze a more powerful MEC that possesses computational power ten times superior to that of an EED, representing a high-performance edge server, to further stress-test EEC's capabilities.

While practical MEC implementations suffer from I/O interference losses between VMs \cite{lit13}, our analysis conservatively assumes lossless parallelization where computational power is perfectly divided, presenting a best-case scenario for MEC performance.
The MEC communication delay is computed similarly to that of the EEDs, while the average computation delay at the MEC is modeled as an exponential random variable with execution rates of $5\mu_f$ and $10\mu_f$ for the 5x and 10x power ratios, respectively, where $\mu_f = 0.007$. This establishes a true parallel-vs-parallel comparison: our distributed EEC architecture against a centralized but parallel MEC system.

Fig. \ref{results_MEC_and_EEDs_time} illustrates the average task response delay as a function of segmentation count (\(n\)) for EEC alongside MEC configurations. The results demonstrate that when (\(\nu_{r_{MEC}} = \nu_r\)), EEC achieves significantly superior performance compared to the practical 5x MEC system, despite our ideal modeling of MEC capabilities. Remarkably, even when challenged by the more powerful 10x MEC configuration, EEC with optimal task segmentation (\(n = 13\)) maintains lower average task response delay. This optimal operating point represents a crucial balance between communication overhead and computational parallelism: below this threshold, EED computational resources remain underutilized, while beyond it, communication costs increase unnecessarily.

Furthermore, the analysis reveals MEC's inherent vulnerability to requester congestion and underscores EEC's core advantages stemming from distributed spatial parallelism. For the 5x MEC case, we observe three distinct performance regimes: minimal delay under no congestion (\(\nu_{r_{MEC}} = 0\)), where the server dedicates full resources to a single task; significant degradation under moderate congestion (\(\nu_{r_{MEC}} = \nu_r\)); and nearly doubled delay under high congestion (\(\nu_{r_{MEC}} = 2\nu_r\)). This pronounced sensitivity to user load contrasts with EEC's congestion-resilient architecture, which effectively leverages underutilized edge resources to enable scalable, low-latency computation while avoiding the single-point congestion bottlenecks that afflict centralized MEC systems.

\subsection{Advanced Model Results}

Here, we conduct a sensitivity analysis of the advanced system model to investigate how key parameters affect performance and to derive system-level design insights.

\begin{figure}[t]
\centering
\begin{tikzpicture}[scale=0.7, transform shape,font=\Large]

\definecolor{color0}{rgb}{0.12156862745098,0.466666666666667,0.705882352941177}
\definecolor{color1}{rgb}{0.00000,0.44700,0.74100}%
\definecolor{color2}{rgb}{0.85000,0.32500,0.09800}%
\definecolor{color3}{rgb}{0.92900,0.69400,0.12500}%
\definecolor{color4}{rgb}{0.49400,0.18400,0.55600}%
\definecolor{color5}{rgb}{0.46600,0.67400,0.18800}%
\definecolor{color6}{rgb}{0.30100,0.74500,0.93300}%
\definecolor{color7}{rgb}{0.63500,0.07800,0.18400}%
\begin{axis}[
width=3.8in,
height=3.2in,
legend cell align={left},
legend style={fill opacity=0.8, draw opacity=1, text opacity=1, draw=white!80!black},
tick align=outside,
tick pos=left,
xlabel={Number of Task Segments ($n$)},
xmin=1, xmax=15,
xtick style={color=black},
ylabel={Average Response Delay (sec)},
ymin=0, ymax=200,
ytick style={color=black},
xmajorgrids,
ymajorgrids,
xtick={0,1,2,...,15},
]
\addplot [line width=0.6mm, color=color1]
table {%
1 201.008990101807
2 152.53151906445
3 126.296204796973
4 109.800898275157
5 98.5462313922673
6 90.4771829796481
7 84.5090686456753
8 80.0102920054823
9 76.5858587679374
10 73.9749367595498
11 71.9977275169447
12 70.5258851964957
13 69.4650848185437
14 68.7442653183958
15 68.3087363062301
};
\addlegendentry{$\mu_f$ = 0.005}
\addplot [line width=0.6mm, color=color2]
table {%
1 101.008990101807
2 77.5263930564363
3 65.1805635974865
4 57.7177642476824
5 52.8889098079606
6 49.6671425389506
7 47.5106306070376
8 46.1039446733109
9 45.249820232226
10 44.8179215031979
11 44.7182488492204
12 44.8863145306814
13 45.2743940400913
14 45.8461202425456
15 46.5730186137017
};
\addlegendentry{$\mu_f$ = 0.01}
\addplot [line width=0.6mm, color=color3]
table {%
1 21.0089901018071
2 17.4887965503071
3 16.2382839277134
4 15.9964912303767
5 16.3091765273252
6 16.9641800699568
7 17.8436930276652
8 18.8736911865573
9 20.0038589615685
10 21.1987060734102
11 22.433053358866
12 23.6892679210495
13 24.9552751598824
14 26.2230254613499
15 27.487313164087
};
\addlegendentry{$\mu_f$ = 0.05}
\addplot [line width=0.6mm, color=color4]
table {%
1 11.0089901018071
2 9.94895864242751
3 10.0461779712739
4 10.6529045652378
5 11.5269912981182
6 12.5488439360113
7 13.6505938535325
8 14.7920412930148
9 15.9496877487507
10 17.1103441244605
11 18.2670226021952
12 19.4163233777514
13 20.5568462285148
14 21.6882666186186
15 22.8108151610967
};
\addlegendentry{$\mu_f$ = 0.1}
\addplot [semithick, red, mark=*, mark size=3, mark options={solid}, only marks, forget plot]
table {%
15 68.3087363062301
};
\addplot [semithick, red, mark=*, mark size=3, mark options={solid}, only marks, forget plot]
table {%
11 44.7182488492204
};
\addplot [semithick, red, mark=*, mark size=3, mark options={solid}, only marks, forget plot]
table {%
4 15.9964912303767
};
\addplot [semithick, red, mark=*, mark size=3, mark options={solid}, only marks, forget plot]
table {%
2 9.94895864242751
};
\addplot [dashed, line width=0.6mm, color1]
table {%
1 201.967226014597
2 154.899088457724
3 130.07834704215
4 114.999155887753
5 105.165319127845
6 98.5256710821482
7 93.9994989990419
8 90.9590930024938
9 89.0131106240253
10 87.903979161649
11 87.4546328799319
12 87.5388163980744
13 88.063556890275
14 88.9583332080801
15 90.1681349870381
};
\addplot [dashed, line width=0.6mm, color2]
table {%
1 101.967226014597
2 79.8808302397817
3 68.949366523514
4 62.9148292338717
5 59.5331844642383
6 57.7830701729495
7 57.1272043017555
8 57.2530721042142
9 57.9642129873228
10 59.1287167486124
11 60.652629667416
12 62.4654071007482
13 64.5116804198878
14 66.7465193997004
15 69.1326776426082
};
\addplot [dashed, line width=0.6mm, color3]
table {%
1 21.9672260145968
2 19.7563743410967
3 19.8456797280603
4 20.9509222303898
5 22.5964675041823
6 24.5478517671524
7 26.6703734430475
8 28.8823072177015
9 31.1343994706208
10 33.3980406306171
11 35.6574632126579
12 37.9045773797584
13 40.1357061649574
14 42.3496209627018
15 44.5464030309196
};
\addplot [dashed, line width=0.6mm, color4]
table {%
1 11.9672260145968
2 12.1403377039498
3 13.4668185309305
4 15.2619901392892
5 17.2645220496791
6 19.3528706084222
7 21.4692549331109
8 23.5876434000517
9 25.6974159276637
10 27.7950332733434
11 29.8800210069054
12 31.9531301057028
13 34.0155245755247
14 36.0684344967352
15 38.1130162870166
};
\addplot [semithick, red, mark=*, mark size=3, mark options={solid}, only marks, forget plot]
table {%
11 87.4546328799319
};
\addplot [semithick, red, mark=*, mark size=3, mark options={solid}, only marks, forget plot]
table {%
7 57.1272043017555
};
\addplot [semithick, red, mark=*, mark size=3, mark options={solid}, only marks, forget plot]
table {%
2 19.7563743410967
};
\addplot [semithick, red, mark=*, mark size=3, mark options={solid}, only marks, forget plot]
table {%
1 11.9672260145968
};
\end{axis}
\end{tikzpicture}
\caption{Average response delay of random (dashed) versus ordered (solid) selection of devices under varying $\mu_f$.}
\label{results-lambda-f-with-order}
\end{figure}
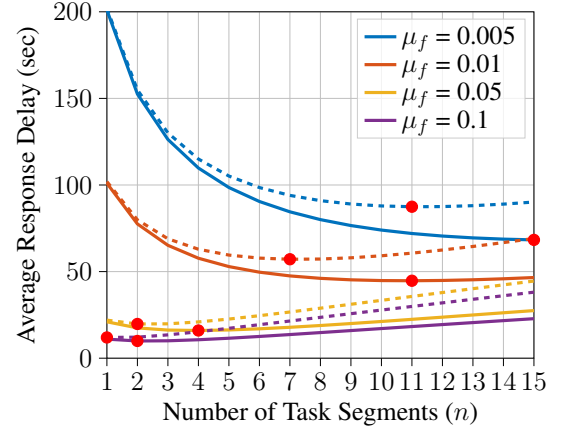

Fig. \ref{results-lambda-f-with-order} plots the average response delay across different task execution rates ($\mu_f$), comparing random EED selection (dashed lines) with ordered selection (solid lines).  Across all $\mu_f$ values, ordered selection achieves a lower average delay and shifts the delay-minimizing segmentation level to a larger optimal $n$ (red markers). This gain, reaching up to about 22\% delay reduction, is primarily driven by shorter offloading links, which increase the offloading success probability and reduce retransmissions, thereby lowering the effective communication cost and allowing the system to exploit parallelism more efficiently. When combined with the low-overhead location acquisition methods discussed in Section \ref{lo_aw_offla}, these results support the practicality of location-aware selection for improving EEC performance.

\begin{figure*}[t]
\centering
\begin{minipage}[c]{0.45\linewidth}
    \centering
\begin{tikzpicture}[scale=0.7, transform shape,font=\Large]

\definecolor{color0}{rgb}{0.12156862745098,0.466666666666667,0.705882352941177}
\definecolor{color1}{rgb}{0.00000,0.44700,0.74100}%
\definecolor{color2}{rgb}{0.85000,0.32500,0.09800}%
\definecolor{color3}{rgb}{0.92900,0.69400,0.12500}%
\definecolor{color4}{rgb}{0.49400,0.18400,0.55600}%
\definecolor{color5}{rgb}{0.46600,0.67400,0.18800}%
\definecolor{color6}{rgb}{0.30100,0.74500,0.93300}%
\definecolor{color7}{rgb}{0.63500,0.07800,0.18400}%
\begin{axis}[
width=3.8in,
height=3.2in,
legend cell align={left},
legend style={fill opacity=0.8, draw opacity=1, text opacity=1, draw=white!80!black},
tick align=outside,
tick pos=left,
xlabel={Number of Task Segments ($n$)},
xmin=1, xmax=17,
xtick style={color=black},
ylabel={Average Response Delay (sec)},
ymin=0, ymax=200,
ytick style={color=black},
xmajorgrids,
ymajorgrids,
xtick={1,3,5,...,17},
]
\addplot [line width=0.6mm, color=color1]
table {%
1 201.008988450767
2 152.531619068284
3 126.297011769653
4 109.803699309592
5 98.5532819574995
6 90.4919548337296
7 84.5364827562598
8 80.0569487131429
9 76.6603338435683
10 74.0882334051768
11 72.1640564215702
12 70.7639768835026
13 69.8002544422471
14 69.2114486677741
15 68.9567193474887
16 69.0127359790757
17 69.3724710447587
18 70.0456470875698
19 71.0610000676621
};
\addlegendentry{$\mu_f$ = 0.005}
\addplot [line width=0.6mm, color=color2]
table {%
1 101.008988450767
2 77.5265870530152
3 65.1821260398706
4 57.7231847460489
5 52.9024736321322
6 49.6953076861997
7 47.5623192580963
8 46.1907714540262
9 45.3863299435373
10 45.0219488013329
11 45.0115766679065
12 45.2958567236614
13 45.8341638913363
14 46.600261971959
15 47.5801829808406
16 48.7715890547046
17 50.1842529634611
18 51.8415309700017
19 53.7828395322479
};
\addlegendentry{$\mu_f$ = 0.01}
\addplot [line width=0.6mm, color=color3]
table {%
1 21.0089884507671
2 17.4896799444485
3 16.244566749284
4 16.0170342586899
5 16.3571246255691
6 17.0559520918041
7 17.9974794853023
8 19.1079969802094
9 20.3365402808251
10 21.6467621537046
11 23.0132543694155
12 24.4196018733434
13 25.8571773868577
14 27.3243701868767
15 28.8261690551528
16 30.3740848944476
17 31.9864480115916
18 33.689195093067
19 35.5173759879223
};
\addlegendentry{$\mu_f$ = 0.05}
\addplot [line width=0.6mm, color=color4]
table {%
1 11.0089884507671
2 9.95057255610675
3 10.0563202885025
4 10.6834132857648
5 11.5926850626421
6 12.6652298784143
7 13.8323697525494
8 15.0526506162153
9 16.3016328652515
10 17.5659149589358
11 18.8393039757648
12 20.1205422570366
13 21.4122106397179
14 22.7204493349273
15 24.055205341293
16 25.4308327203688
17 26.867009734174
18 28.3900755770847
19 30.0350254404863
};
\addlegendentry{$\mu_f$ = 0.1}
\addplot [semithick, red, mark=*, mark size=3, mark options={solid}, only marks, forget plot]
table {%
15 68.3087363062301
};
\addplot [semithick, red, mark=*, mark size=3, mark options={solid}, only marks, forget plot]
table {%
11 44.7182488492204
};
\addplot [semithick, red, mark=*, mark size=3, mark options={solid}, only marks, forget plot]
table {%
4 15.9964912303767
};
\addplot [semithick, red, mark=*, mark size=3, mark options={solid}, only marks, forget plot]
table {%
2 9.94895864242751
};
\addplot [dashed, line width=0.6mm, color1]
table {%
1 336.620680186931
2 285.180322992517
3 242.744420867077
4 212.348252260252
5 190.293257161066
6 173.829658800683
7 161.233627933956
8 151.417092124124
9 143.669683869886
10 137.511058188124
11 132.606423070036
12 128.716980466865
13 125.670086463533
14 123.340979943515
15 121.641612448277
16 120.514113324023
17 119.927604227705
};
\addplot [dashed, line width=0.6mm, color2]
table {%
1 185.580938612364
2 150.226713879031
3 126.510893278056
4 111.088921733864
5 100.588768964621
6 93.1951754287126
7 87.8966777061496
8 84.0914796425699
9 81.4003455226771
10 79.5729996143869
11 78.4383398452359
12 77.8766316048772
13 77.8035366888404
14 78.1609670133352
15 78.9121452893949
16 80.0394511887847
17 81.544283484348
};
\addplot [dashed, line width=0.6mm, color3]
table {%
1 40.8662306522036
2 33.0122018773198
3 29.4918082914123
4 27.9895233075779
5 27.6139149825816
6 27.9468838973397
7 28.7555797898016
8 29.8919874701709
9 31.2536541156629
10 32.7671851162511
11 34.3805415720847
12 36.0586823406242
13 37.780555741267
14 39.5368372161389
15 41.3282670370395
16 43.1645466709728
17 45.0637908179335
};
\addplot [dashed, line width=0.6mm, color4]
table {%
1 21.03760007082
2 17.9149370961077
3 17.0883668515656
4 17.2975083719612
5 18.062667263418
6 19.1449349521725
7 20.4063641474584
8 21.7636529844031
9 23.167496153841
10 24.5904062762989
11 26.0187500274496
12 27.4477655358506
13 28.8787865667855
14 30.3179689458989
15 31.7759564392477
16 33.2681297784043
17 34.8152953039477
};
\addplot [semithick, red, mark=*, mark size=3, mark options={solid}, only marks, forget plot]
table {%
17 119.927604227705
};
\addplot [semithick, red, mark=*, mark size=3, mark options={solid}, only marks, forget plot]
table {%
13 77.8035366888404
};
\addplot [semithick, red, mark=*, mark size=3, mark options={solid}, only marks, forget plot]
table {%
5 27.6139149825816
};
\addplot [semithick, red, mark=*, mark size=3, mark options={solid}, only marks, forget plot]
table {%
3 17.0883668515656
};
\end{axis}
\end{tikzpicture}
    \caption{ Average task response delay, with EED failure (dashed) and without EED failure (solid), for varying task execution rates ($\mu_f$).}
    \label{fail_1}
\end{minipage}
\hspace*{5mm}
\begin{minipage}[c]{0.45\linewidth}
    \centering
\begin{tikzpicture}[scale=0.7, transform shape,font=\Large]

\definecolor{color0}{rgb}{0.12156862745098,0.466666666666667,0.705882352941177}
\definecolor{color1}{rgb}{0.00000,0.44700,0.74100}%
\definecolor{color2}{rgb}{0.85000,0.32500,0.09800}%
\definecolor{color3}{rgb}{0.92900,0.69400,0.12500}%
\definecolor{color4}{rgb}{0.49400,0.18400,0.55600}%
\definecolor{color5}{rgb}{0.46600,0.67400,0.18800}%
\definecolor{color6}{rgb}{0.30100,0.74500,0.93300}%
\definecolor{color7}{rgb}{0.63500,0.07800,0.18400}%
\begin{axis}[
width=3.8in,
height=3.2in,
legend cell align={left},
legend style={at={(0.9,0.55)}, font=\large, fill opacity=0.8, draw opacity=1, text opacity=1, draw=white!80!black},
tick align=outside,
tick pos=left,
xlabel={Number of Task Segments ($n$)},
xmin=1, xmax=17,
xtick style={color=black},
ylabel={Average Response Delay (sec)},
ymin=0, ymax=200,
ytick style={color=black},
xmajorgrids,
ymajorgrids,
]
\addplot[dashed,
  line width=0.8mm,
  color=color1,
  mark size=2pt,
  mark repeat=1, 
]
table {%
1 99.1930348331605
2 81.946794204932
3 79.6295949127609
4 98.9440815376948
5 169.721429314715
6 437.163906258704
};
\addlegendentry{$\nu_w = 5 \times 10^{-5}$}
\addplot[dashed,
  line width=0.8mm,
  color=color2,
  mark size=2pt,
  mark repeat=1, 
]
table {%
1 98.4350614861812
2 78.6984803390864
3 68.8180271485472
4 65.7789019682192
5 70.1201080457109
6 85.8686291706936
7 122.68942672516
8 213.259132661576
};
\addlegendentry{$\nu_w=0.0001$ }
\addplot[dashed,
  line width=0.8mm,
  color=color4,
]
table {%
1 98.1980746330094
2 77.7401114329836
3 66.161583417178
4 59.2767522449326
5 55.0211468941131
6 52.4053035254994
7 50.9005524317831
8 50.197502412881
9 50.1009045469313
10 50.4789828020191
11 51.2370674986563
12 52.3031864907215
13 53.6200191917846
14 55.1404131509608
15 56.8249295273782
16 58.6405260036283
17 60.5598546097105
18 62.5608828128583
};
\addlegendentry{$\nu_w=0.001$ }
\addplot [semithick, red, mark=*, mark size=3, mark options={solid}, only marks]
table {%
3 79.6295949127609
};
\addplot [semithick, red, mark=*, mark size=3, mark options={solid}, only marks]
table {%
4 65.7789019682192
};
\addplot [semithick, red, mark=*, mark size=3, mark options={solid}, only marks]
table {%
9 50.1009045469313
};

\draw[red, line width=1pt]
  (axis cs:5.3,175) ellipse [x radius=3.0, y radius=12];

\draw[->, red, line width=1pt]
  (axis cs:7.6,170) -- (axis cs:10.4,145);

\node[red, font=\Large, anchor=west, align=left]
  at (axis cs:10.4,145) {Congestion at\\workers};

\end{axis}

\end{tikzpicture}
    \caption{ Average response delay over different $\nu_w$ values under the ordered offloading while considering failure.}
    \label{fail_2}
\end{minipage}
\end{figure*}

In Fig.~\ref{fail_1}, as expected, incorporating device failures increases the average task response delay. It also shifts the optimal number of task segments $n$ to a larger value. This behavior stems from the coupling between the failure rate $\gamma$ and the execution rate $\mu_f$: when the task is divided into more segments, each worker handles a smaller portion of the task, which effectively reduces the per-segment failure rate $\gamma_n$ and lowers the probability of failure events. However, this trend persists only up to a certain segmentation level. Beyond that point, further splitting the task requires allocating more devices, which significantly increases the offloading time; the resulting communication cost eventually dominates the reduction in failure probability and causes the average task response delay to rise. This figure therefore highlights the sensitivity of both the delay and the optimal segmentation level to the failure parameter $\gamma$ and underscores the importance of jointly accounting for reliability and communication overhead when selecting $n$.

To examine congestion due to worker scarcity, we conduct an additional experiment to evaluate the impact of the worker intensity $\nu_w$ on the average task response delay. As shown in Fig.~\ref{fail_2}, lower values of $\nu_w$ lead to significantly higher delays and a smaller optimal number of task segments. This behavior is primarily due to the limited number of available LoS EEDs, which induces congestion as multiple requesters compete for a scarce set of workers. In such cases, the few accessible workers are often located farther from the requester, resulting in weaker D2D links and a reduced offloading success probability, thereby increasing the time required to allocate each task segment.
The figure further shows that, under low $\nu_w$, the average response delay rises sharply beyond a certain segmentation level and continues to increase as $n$ grows, reflecting the growing difficulty of recruiting reliable LoS workers. Consequently, in worker-congested regimes, deviations from the optimal segmentation level incur a much larger delay penalty. This sensitivity to $\nu_w$ highlights the importance of the EEC-MEC collaboration approach proposed in this work: by offloading excess demand to MEC resources when worker intensity is low, the system can effectively mitigate congestion at the EED level and maintain low response delays under constrained worker availability.

\begin{figure*}[t]
\centering
\begin{minipage}[c]{0.45\linewidth}
    \centering
    \includegraphics[width=1\linewidth]{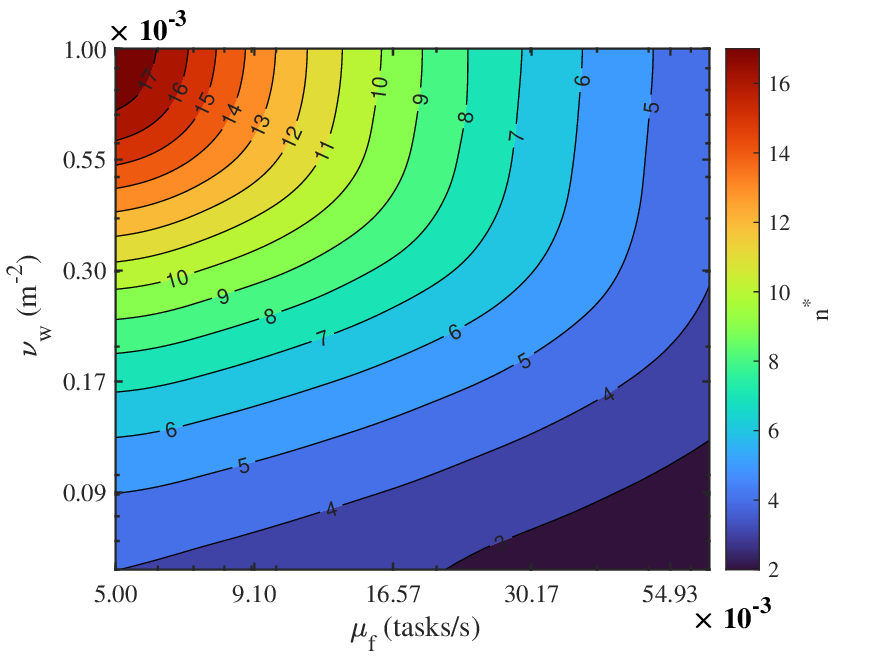}
    \caption{The optimal number of task segments ($n^*$) as a joint function of worker intensity ($\nu_w$) and task execution rate ($\mu_f$).}
    \label{fig:new_contour_map}
\end{minipage}
\hspace*{5mm}
\begin{minipage}[c]{0.45\linewidth}
\centering
\begin{tikzpicture}[scale=0.7, transform shape,font=\Large]

\definecolor{color0}{rgb}{0.12156862745098,0.466666666666667,0.705882352941177}
\definecolor{color1}{rgb}{0.00000,0.44700,0.74100}%
\definecolor{color2}{rgb}{0.85000,0.32500,0.09800}%
\definecolor{color3}{rgb}{0.92900,0.69400,0.12500}%
\definecolor{color4}{rgb}{0.49400,0.18400,0.55600}%
\definecolor{color5}{rgb}{0.46600,0.67400,0.18800}%
\definecolor{color6}{rgb}{0.30100,0.74500,0.93300}%
\definecolor{color7}{rgb}{0.63500,0.07800,0.18400}%
\begin{axis}[
width=3.8in,
height=3.2in,
legend cell align={left},
legend style={at={(0.9,0.29)}, font=\large, fill opacity=0.8, draw opacity=1, text opacity=1, draw=white!80!black},
tick align=outside,
tick pos=left,
xlabel={Number of Task Segments ($n$)},
xmin=1, xmax=18,
xtick style={color=black},
ylabel={Task Completion Probability},
ymin=0.5, ymax=1,
ytick style={color=black},
xmajorgrids,
ymajorgrids,
]
\addplot[
  line width=0.6mm,
  color=color1,
  mark=square,
  mark size=2pt,
  mark repeat=1, 
]
table {%
1  0.5
2  0.64
3  0.729
4  0.784664934567355
5  0.821927106759351
6  0.848407814786616
7  0.868125533246719
8  0.883350254966405
9  0.897
10 0.9052
11 0.9142
12 0.9235
13 0.926
14 0.929
15 0.934
16 0.9383
17 0.9429
18 0.9463
};
\addlegendentry{$l=1$ }
\addplot[
  line width=0.6mm,
  color=color2,
  mark=triangle,
  mark size=2pt,
  mark repeat=1, 
]
table {%
1  0.666666666666667
2  0.790123456790124
3  0.850269718617874
4  0.884187057991215
5  0.905730809829916
6  0.920572119871645
7  0.931399858600204
8  0.939641251845746
9  0.9443
10 0.9542
11 0.9564
12 0.9587
13 0.9622
14 0.9667
15 0.9685
16 0.9696
17 0.973
18 0.975
};
\addlegendentry{$l=2$ }
\addplot[
  line width=0.6mm,
  color=color4,
  mark=*,
  mark size=2pt,
  mark repeat=1, 
]
table {%
1  0.833333333333333
2  0.90702947845805
3  0.936190104380702
4  0.951524275217153
5  0.960942362645598
6  0.967305331327279
7  0.971889389591644
8  0.975347931627432
9  0.9773
10 0.9793
11 0.9821
12 0.9833
13 0.985
14 0.9872
15 0.9877
16 0.9882
17 0.989
18 0.99
};
\addlegendentry{$l=5$ }
\end{axis}

\end{tikzpicture}
    \caption{Task completion probability over varying EEDs reliability Parameters}
    \label{comp_prob}
\end{minipage}
\end{figure*}

To guide system design, Fig.~\ref{fig:new_contour_map} plots the contour of the optimal task segmentation, $n^*$, as a joint function of worker intensity ($\nu_w$) and task execution rate ($\mu_f$). This design map reveals three core operational regimes. First, in the high $\nu_w$ and low $\mu_f$ regime, $n^*$ is high. This is because abundant nearby workers minimize the communication overhead of offloading, while the slow task execution rate ensures that the computation speedup from parallelization is significant enough to outweigh these communication costs. Second, under low $\nu_w$, $n^*$ is forced to be low regardless of $\mu_f$, confirming that worker scarcity creates a communication bottleneck that dominates performance. Third, for high $\mu_f$, $n^*$ remains low even with ample workers, as the minimal computation time makes parallelization gains insignificant compared to the fixed offloading cost.
Collectively, this map provides a vital practical tool, showing that optimal operation requires matching the segmentation strategy to the specific environment. Aggressive parallelization is only beneficial in dense networks with computationally intensive tasks, while simpler tasks or sparse networks require minimal segmentation to avoid excessive communication overhead.

Fig.~\ref{comp_prob} illustrates the task completion probability as a function of the EED reliability parameter $l$. As expected, the completion probability decreases as $l$ decreases, reflecting a higher likelihood of worker failure and, consequently, a reduced chance of successful task execution. More importantly, increasing the number of task segments $n$ consistently enhances the completion probability. This improvement arises because smaller segments have shorter execution durations, which reduces the risk that a worker fails before finishing its assigned portion.
However, beyond a certain segmentation threshold, the marginal gains in reliability begin to diminish. As segment sizes become very small, further segmentation provides limited additional benefit while incurring extra communication overhead, which can increase the total task response delay. Hence, achieving very high reliability (e.g., a completion probability of $0.99$) may require operating at a segmentation level $n$ that is larger than the value minimizing the average response delay. Such scenarios often arise in applications with stringent reliability requirements, where the system designer may deliberately choose a higher $n$ to ensure task completion, even at the expense of increased latency.

These results highlight the value of considering task completion probability as a reliability metric alongside the average task response delay. Together, these metrics provide a more complete perspective for selecting a segmentation level $n$ that balances latency and reliability for a given application. It is also worth noting that the system congestion level plays a critical role in this trade-off. As discussed earlier, while increasing $n$ generally improves reliability, the associated delay penalty is highly sensitive to congestion. For example, in highly congested scenarios with low worker intensity, pursuing very high reliability by increasing $n$ can cause a sharp rise in response delay, as clearly illustrated in Fig.~\ref{fail_2}.

\begin{figure*}[!t]
  \centering
  \subfloat[Low task execution rate $\mu_f=0.002$\label{fig:subfig-a}]{
    \begin{minipage}[b]{0.48\linewidth}\centering
      \scalebox{0.90}{

\begin{tikzpicture}[scale=0.85, transform shape,font=\Large]

\definecolor{color0}{rgb}{0.12156862745098,0.466666666666667,0.705882352941177}

\definecolor{color1}{rgb}{0.00000,0.44700,0.74100}%
\definecolor{color2}{rgb}{0.85000,0.32500,0.09800}%
\definecolor{color3}{rgb}{0.92900,0.69400,0.12500}%
\definecolor{color4}{rgb}{0.49400,0.18400,0.55600}%
\definecolor{color5}{rgb}{0.46600,0.67400,0.18800}%
\definecolor{color6}{rgb}{0.30100,0.74500,0.93300}%
\definecolor{color7}{rgb}{0.63500,0.07800,0.18400}%

\begin{axis}[
width=3in,
height=2.6in,
legend cell align={left},
legend style={font=\footnotesize, fill opacity=0.8, draw opacity=1, text opacity=1, draw=white!80!black},
tick align=outside,
tick pos=left,
x grid style={white!69.0196078431373!black},
xlabel={\normalsize{ Tasks Offloading Bias Towards EEC ($\alpha$)}},
xmin=0, xmax=1,
xtick style={color=black},
y grid style={white!69.0196078431373!black},
ylabel={\normalsize{Average Response Delay (sec)}},
ymin=0, ymax=600,
ytick style={color=black},
xtick={0, 0.2, ..., 1},
xmajorgrids,
ymajorgrids
]
\addplot [line width=0.6mm, color1]
table {%
0 134.783047212379
0.1 252.066881576197
0.2 302.623419870897
0.3 335.166485670186
0.4 358.594639792094
0.5 391.389020663817
0.6 396.450711653525
0.7 402.182418040047
0.8 408.566018885414
0.9 415.583390469717
1 423.216379993177
};
\addlegendentry{EEC-Only Requesters}
\addplot [dashed, line width=0.9mm, color4]
table {%
0 416.159265358979
0.1 384.743338823081
0.2 353.327412287183
0.3 321.911485751286
0.4 290.495559215388
0.5 259.07963267949
0.6 227.663706143592
0.7 196.247779607694
0.8 164.831853071796
0.9 133.415926535898
1 102
};
\addlegendentry{MEC-Only Requesters }
\addplot [solid, line width=0.4mm, color5]
table {%
0 416.159265358979
0.1 371.475693098393
0.2 343.186613803926
0.3 325.887985726956
0.4 317.73519144607
0.5 325.234326671653
0.6 328.935909449551
0.7 340.402026510341
0.8 359.81918572269
0.9 387.366644076335
1 423.216379993177
};
\addlegendentry{Combined System}
\end{axis}
\end{tikzpicture}


}
    \end{minipage}
  }\hfill
  \subfloat[Low worker intensity $\nu_w/4$, $\mu_f = 0.02$\label{fig:subfig-b}]{
    \begin{minipage}[b]{0.48\linewidth}\centering
      \scalebox{0.90}{

\begin{tikzpicture}[scale=0.85, transform shape,font=\Large]

\definecolor{color0}{rgb}{0.12156862745098,0.466666666666667,0.705882352941177}

\definecolor{color1}{rgb}{0.00000,0.44700,0.74100}%
\definecolor{color2}{rgb}{0.85000,0.32500,0.09800}%
\definecolor{color3}{rgb}{0.92900,0.69400,0.12500}%
\definecolor{color4}{rgb}{0.49400,0.18400,0.55600}%
\definecolor{color5}{rgb}{0.46600,0.67400,0.18800}%
\definecolor{color6}{rgb}{0.30100,0.74500,0.93300}%
\definecolor{color7}{rgb}{0.63500,0.07800,0.18400}%

\begin{axis}[
width=3in,
height=2.6in,
legend cell align={left},
legend style={font=\footnotesize,fill opacity=0.8, draw opacity=1, text opacity=1, draw=white!80!black},
tick align=outside,
tick pos=left,
x grid style={white!69.0196078431373!black},
xlabel={\normalsize{ Tasks Offloading Bias Towards EEC ($\alpha$)}},
xmin=0, xmax=1,
xtick style={color=black},
y grid style={white!69.0196078431373!black},
ylabel={\normalsize{Average Response Delay (sec)}},
ymin=0, ymax=100,
ytick style={color=black},
xtick={0, 0.2, ..., 1},
xmajorgrids,
ymajorgrids
]
\addplot [line width=0.6mm, color1]
table {%
0 32.1002339800581
0.1 42.8666152792519
0.2 47.564225316581
0.3 52.5175181221595
0.4 53.0844551278437
0.5 53.6553175110807
0.6 54.2282746392819
0.7 54.8024798611265
0.8 55.3774882492936
0.9 55.9530437588186
1 56.5289886668121
};
\addlegendentry{EEC-Only Requesters}
\addplot [dashed, line width=0.9mm, color4]
table {%
0 43.4159265358979
0.1 40.2743338823081
0.2 37.1327412287183
0.3 33.9911485751285
0.4 30.8495559215388
0.5 27.707963267949
0.6 24.5663706143592
0.7 21.4247779607694
0.8 18.2831853071796
0.9 15.1415926535898
1 12
};
\addlegendentry{MEC-Only Requesters }
\addplot [solid, line width=0.4mm, color5]
table {%
0 43.4159265358979
0.1 40.5335620220025
0.2 39.2190380462909
0.3 39.5490594392378
0.4 39.7435156040607
0.5 40.6816403895148
0.6 42.3635130293128
0.7 44.7891692910194
0.8 47.9586276608708
0.9 51.8718986482957
1 56.5289886668121
};
\addlegendentry{Combined System}
\end{axis}
\end{tikzpicture}


}
    \end{minipage}
  }

  \vspace{0.5ex}

  \subfloat[Low requester intensity $\nu_r/4$, $\mu_f = 0.02$\label{fig:subfig-c}]{
    \begin{minipage}[b]{0.48\linewidth}\centering
      \scalebox{0.90}{

\begin{tikzpicture}[scale=0.85, transform shape,font=\Large]

\definecolor{color0}{rgb}{0.12156862745098,0.466666666666667,0.705882352941177}

\definecolor{color1}{rgb}{0.00000,0.44700,0.74100}%
\definecolor{color2}{rgb}{0.85000,0.32500,0.09800}%
\definecolor{color3}{rgb}{0.92900,0.69400,0.12500}%
\definecolor{color4}{rgb}{0.49400,0.18400,0.55600}%
\definecolor{color5}{rgb}{0.46600,0.67400,0.18800}%
\definecolor{color6}{rgb}{0.30100,0.74500,0.93300}%
\definecolor{color7}{rgb}{0.63500,0.07800,0.18400}%

\begin{axis}[
width=3in,
height=2.6in,
legend cell align={left},
legend style={font=\footnotesize,fill opacity=0.8, draw opacity=1, text opacity=1, draw=white!80!black},
tick align=outside,
tick pos=left,
x grid style={white!69.0196078431373!black},
xlabel={\normalsize{ Tasks Offloading Bias Towards EEC ($\alpha$)}},
xmin=0, xmax=1,
xtick style={color=black},
y grid style={white!69.0196078431373!black},
ylabel={\normalsize{Average Response Delay (sec)}},
ymin=0, ymax=50,
ytick style={color=black},
xtick={0, 0.2, ..., 1},
xmajorgrids,
ymajorgrids
]
\addplot [line width=0.6mm, color1]
table {%
0 28.8219209119833
0.1 28.8391311778787
0.2 28.8607769008043
0.3 28.8941057766056
0.4 28.9492494933521
0.5 29.0370530489037
0.6 29.1674242908561
0.7 29.3487763624424
0.8 29.4803108448974
0.9 29.6308196084987
1 29.8170722723597
};
\addlegendentry{EEC-Only Requesters}
\addplot [dashed, line width=0.9mm, color4]
table {%
0 19.8539816339745
0.1 19.068583470577
0.2 18.2831853071796
0.3 17.4977871437821
0.4 16.7123889803847
0.5 15.9269908169872
0.6 15.1415926535898
0.7 14.3561944901923
0.8 13.5707963267949
0.9 12.7853981633974
1 12
};
\addlegendentry{MEC-Only Requesters }
\addplot [solid, line width=0.4mm, color5]
table {%
0 19.8539816339745
0.1 20.0456382413072
0.2 20.3987036259045
0.3 20.9166827336292
0.4 21.6071331855716
0.5 22.4820219329455
0.6 23.5570916359496
0.7 24.8510018007674
0.8 26.2984079412769
0.9 27.9462774639886
1 29.8170722723597
};
\addlegendentry{Combined System}
\end{axis}
\end{tikzpicture}


}
    \end{minipage}
  }\hfill
  \subfloat[High requester intensity $4\,\nu_r$, $\mu_f = 0.02$\label{fig:subfig-d}]{
    \begin{minipage}[b]{0.48\linewidth}\centering
      \scalebox{0.90}{
\begin{tikzpicture}[scale=0.85, transform shape,font=\Large]

\definecolor{color0}{rgb}{0.12156862745098,0.466666666666667,0.705882352941177}

\definecolor{color1}{rgb}{0.00000,0.44700,0.74100}%
\definecolor{color2}{rgb}{0.85000,0.32500,0.09800}%
\definecolor{color3}{rgb}{0.92900,0.69400,0.12500}%
\definecolor{color4}{rgb}{0.49400,0.18400,0.55600}%
\definecolor{color5}{rgb}{0.46600,0.67400,0.18800}%
\definecolor{color6}{rgb}{0.30100,0.74500,0.93300}%
\definecolor{color7}{rgb}{0.63500,0.07800,0.18400}%
\begin{axis}[
width=3in,
height=2.6in,
legend cell align={left},
legend style={font=\footnotesize,fill opacity=0.8, draw opacity=1, text opacity=1, draw=white!80!black},
tick align=outside,
tick pos=left,
x grid style={white!69.0196078431373!black},
xlabel={\normalsize{ Tasks Offloading Bias Towards EEC ($\alpha$)}},
xmin=0, xmax=1,
xtick style={color=black},
y grid style={white!69.0196078431373!black},
ylabel={\normalsize{Average Response Delay (sec)}},
ymin=0, ymax=150,
ytick style={color=black},
xtick={0, 0.2, ..., 1},
xmajorgrids,
ymajorgrids
]
\addplot [line width=0.6mm, color1]
table {%
0 30.7667058834262
0.1 35.5364590183949
0.2 39.2647697777718
0.3 42.9181232302872
0.4 44.2923359523677
0.5 45.8609931211967
0.6 47.6069846621686
0.7 49.517755661767
0.8 51.5835383968982
0.9 52.8064539385986
1 53.02229965705
};
\addlegendentry{EEC-Only Requesters}
\addplot [dashed, line width=0.9mm, color4]
table {%
0 137.663706143592
0.1 125.097335529233
0.2 112.530964914873
0.3 99.9645943005142
0.4 87.398223686155
0.5 74.8318530717959
0.6 62.2654824574367
0.7 49.6991118430775
0.8 37.1327412287183
0.9 24.5663706143592
1 12
};
\addlegendentry{MEC-Only Requesters }
\addplot [solid, line width=0.4mm, color5]
table {%
0 137.663706143592
0.1 116.141247878149
0.2 97.8777258874531
0.3 82.8506529794461
0.4 70.1558685926401
0.5 60.3464230964963
0.6 53.4703837802758
0.7 49.5721625161602
0.8 48.6933789632622
0.9 49.9824456061747
1 53.02229965705
};
\addlegendentry{Combined System}
\end{axis}

\end{tikzpicture}}
    \end{minipage}
  }

  \caption{Average task response delay vs.\ portion of requesters offloading to \textit{EED} ($\alpha$) under different system parameters.}
  \label{fig:params_alpha}
\end{figure*}
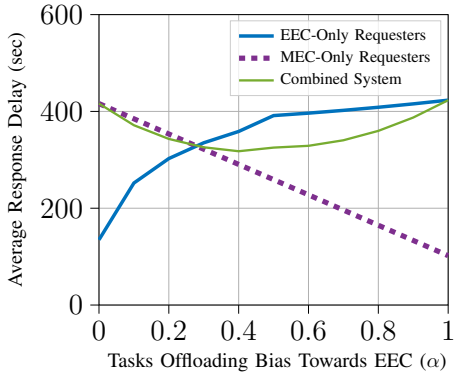
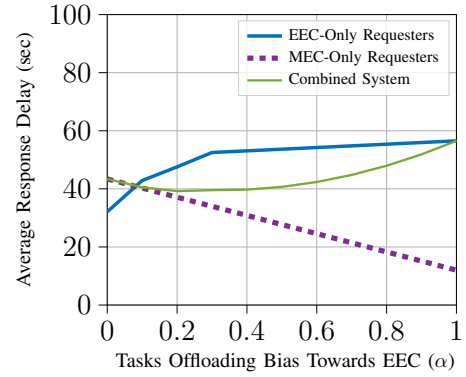
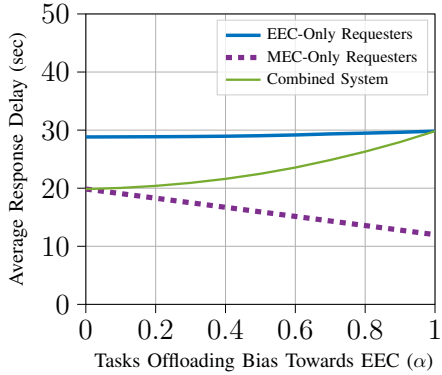
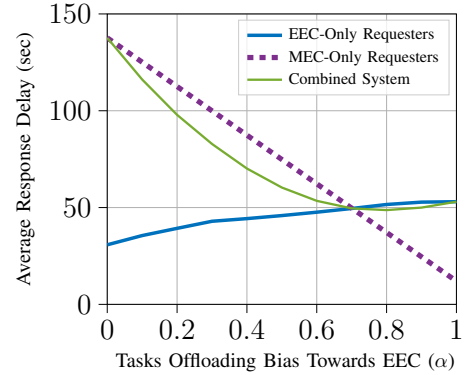

Next, we investigate EEC-MEC collaboration and how the optimal $\alpha$ systematically shifts in response to changing congestion conditions, quantified through the effective idle worker intensity $\nu_{w_{idle}}$. Using the ordered offloading with MEC computational power set to five times that of a single EED, Fig. \ref{fig:params_alpha} depicts the average response delay for: the portion of requesters using EEC exclusively, the portion utilizing MEC exclusively, and the combined system across varying bias factor $\alpha$ values. The combined system represents the overall performance metric we aim to optimize, reflecting the average response delay experienced by any typical requester in the system.
The depicted scenario begins with a single requester utilizing EEC, while the others offload to MEC. Subsequently, as $\alpha$ increases, the situation gradually shifts until one requester exclusively relies on MEC and the rest use EEC.

Fig. \ref{fig:params_alpha}(a) shows the average response delay under low task execution rate, which inherently reduce idle worker density and contribute to system congestion. The low execution rate prolongs task completion times, decreasing the effective idle worker intensity $\nu_{w_{idle}}$ and constraining the available EED pool. This congestion effect reveals an optimal operating point at $\alpha=0.4$, where the system balances EEC utilization against resource scarcity. As $\alpha$ increases from zero, growing EEC usage alleviates MEC congestion, initially reducing combined system delay. However, beyond $\alpha=0.4$, the limited LoS EED availability becomes critically strained, forcing allocations to more distant workers and degrading D2D performance, ultimately increasing the average response delay.
The optimal $\alpha$ thus represents the precise balance that minimizes system-wide delay by avoiding both excessive MEC load (at low $\alpha$) and EEC resource saturation (at high $\alpha$). 

Fig. \ref{fig:params_alpha}(b) examines the impact of low worker intensity, revealing how fundamental resource scarcity shifts the optimal operating point to $\alpha = 0.2$. The critically low worker density dramatically reduces $\nu_{w_{idle}}$ and limits the availability of LoS EEDs. This scarcity forces the allocation of more distant workers, increasing communication delays and retransmissions. The optimal $\alpha$ shift to 0.2 represents the system's adaptation to this constrained environment, where only minimal EEC utilization can be supported without overwhelming the limited worker pool and degrading overall performance.

Fig.~\ref{fig:params_alpha}(c) presents the average response delay under low requester intensity, which significantly increases $\nu_{w_{idle}}$ and reduces system-wide congestion. The optimal operating point at $\alpha = 0$ demonstrates that exclusive MEC offloading minimizes average response delay. In this uncongested regime, MEC's superior computational power provides faster task completion than distributed EEC processing, despite the availability of LoS EEDs. While EEC performance remains stable across $\alpha$ values due to ample worker availability, the absence of requester-level congestion eliminates MEC's scalability limitations, allowing its computational advantage to dominate. This result validates our framework's ability to identify when centralized MEC resources outperform distributed EEC capabilities based on system congestion conditions quantified through $\nu_{w_{idle}}$.

Fig. \ref{fig:params_alpha}(d) presents a high requester intensity scenario, where the optimal operating point shifts dramatically to $\alpha = 0.8$. This significant result demonstrates that extensive EEC utilization becomes the dominant strategy under high congestion, as its distributed nature provides crucial scalability that centralized MEC cannot match when overwhelmed by high requester density.
The optimal $\alpha=0.8$ indicates that 80\% of requesters should be served by EEC resources, leveraging the parallel processing capabilities of EEDs to alleviate MEC congestion. This finding clearly validates our paper's core contribution: EEC emerges as a vital computational paradigm in high-density scenarios, where its distributed architecture and spatial resource pooling overcome the scalability limitations of traditional edge computing.

\section{ Conclusion and Future Work}

This paper presents a novel spatiotemporal framework for EEC in large-scale mmWave networks, integrating SG with an ACTMC to jointly model mmWave D2D offloading and parallel computation, including their temporal overlap. The framework enables a tractable end-to-end evaluation of two key metrics, the average task response delay and the task completion probability, providing a unified view of latency and reliability. Our analysis, validated by Monte Carlo simulations and sensitivity studies, reveals a fundamental communication-computation trade-off. This trade-off yields an optimal task segmentation level that minimizes delay, balancing the benefits of parallelism against the overhead of excessive offloading. This optimum depends critically on operating conditions, such as D2D link quality, computation speed, and worker density, making aggressive parallelization beneficial for computation-heavy tasks with sufficient nearby workers. Furthermore, location-aware EED selection consistently outperforms random selection; by improving offloading success, it pushes the delay-optimal segmentation level higher, enabling greater parallelism gains.

Extending the analysis to practical impairments, we find that EED failures shift the delay-optimal segmentation level higher, as smaller segments enhance resilience against device failure. Conversely, worker scarcity reduces the optimal segmentation level to limit offloading overhead; in such resource-constrained scenarios, deviating from the optimum incurs a severe delay penalty. Furthermore, meeting stringent reliability targets may require operating above the delay-optimal point, underscoring the trade-off between latency and reliability. Finally, regarding EEC-MEC collaboration, we demonstrate that the optimal bias factor adapts to congestion by balancing MEC load against worker scarcity to minimize system delay. Overall, this work provides a rigorous analytical foundation and practical, sensitivity-aware guidelines for EEC system design, enabling informed decisions on task segmentation, offloading strategies, and collaboration across a broad range of operating conditions.

Future work will generalize the proposed spatiotemporal framework to support heterogeneous segment sizes and to incorporate dependent-task workflows with precedence constraints, such as DAG-structured workflows, including synchronization requirements. We plan to explore two complementary directions: (i) a stage-wise abstraction that represents the workflow as an ordered sequence of dependency-constrained stages, where the proposed analysis is applied at the stage level and the end-to-end delay is obtained by composing stage delays; and (ii) a partition-based abstraction that groups dependent subtasks into a small number of macro-tasks executed under precedence constraints, together with corresponding readiness-aware extensions to the delay characterization. In parallel, heterogeneous EED capabilities will be addressed by incorporating selection policies that map segments or workflow components to workers according to their computational power. Finally, we also plan to extend the framework to applications with non-negligible result payloads. This requires augmenting the ACTMC with an explicit result-return phase, and studying how result communication influences segmentation decisions, location-aware offloading benefits, and the MEC-EEC load-balancing bias factor.

\bibliographystyle{IEEEtran}
\bibliography{bibliography.bib}

\begin{IEEEbiography}[{\includegraphics[width=1in,height=1.25in,clip,keepaspectratio]{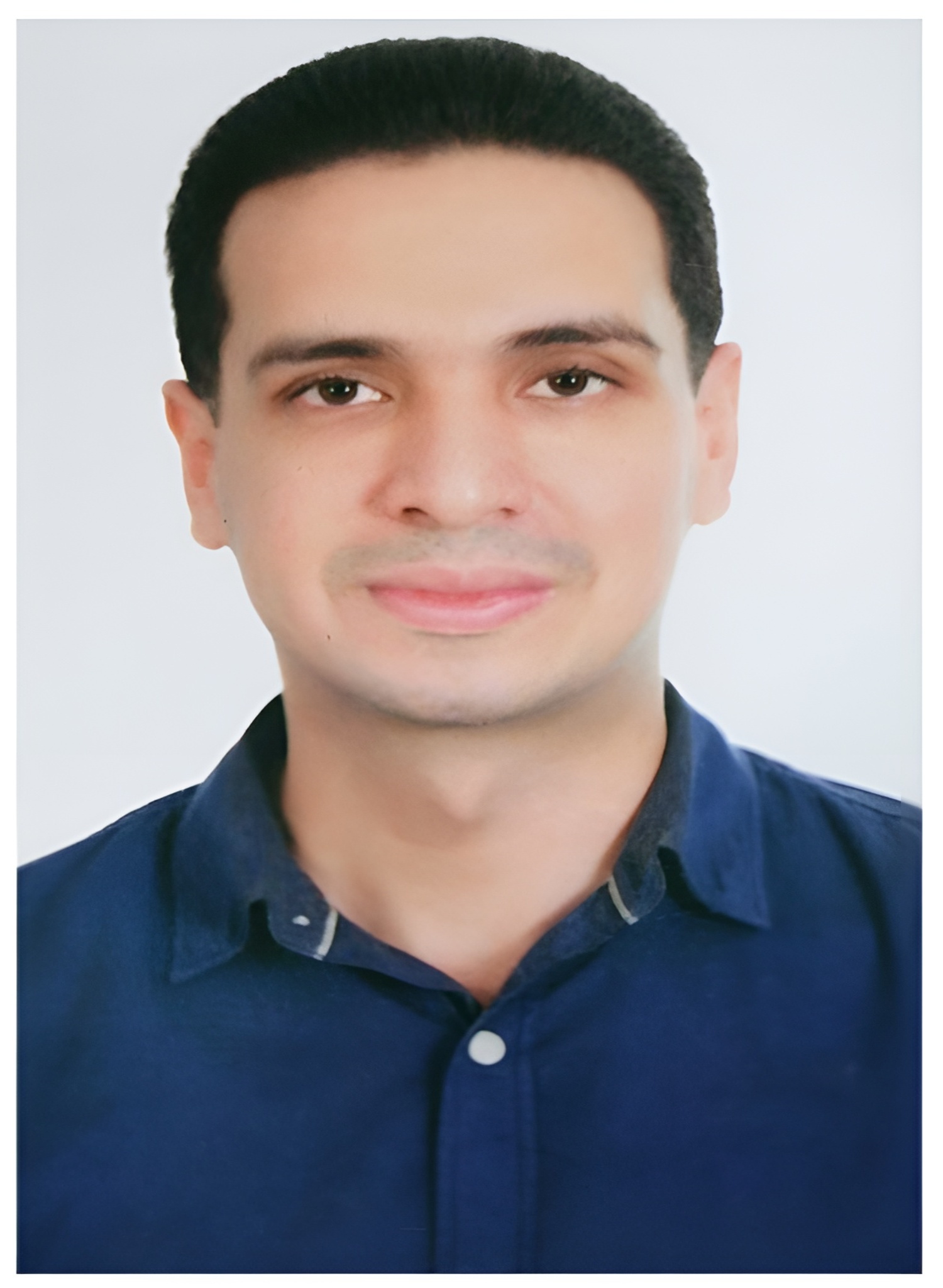}}]{Yasser Nabil}  (Graduate Student Member, IEEE) received the B.Sc. degree in electronics and communications engineering from Alexandria University, Alexandria, Egypt, and the M.Sc. degree in electronics and communications
engineering from Cairo University, Giza, Egypt. He is currently pursuing the Ph.D. degree in electrical and computer engineering at Queen’s University, Kingston, ON, Canada. His research interests include the integration of sensing and communication, wireless communication systems and their statistical modeling, the Internet of Things, non-terrestrial networks, and edge computing. His work has appeared in leading IEEE journals and conferences, including the IEEE Transactions on Wireless Communications, the IEEE Internet of Things Journal, the IEEE Transactions on Vehicular Technology, and IEEE ICC.
\end{IEEEbiography}

\begin{IEEEbiography}[{\includegraphics[width=1in,height=1.25in,clip,keepaspectratio]{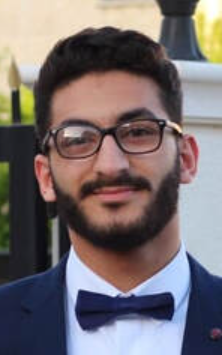}}]{Mahmoud Abdelhadi} holds a Bachelor of Science degree in Computer Science from the Applied Science University in Amman, Jordan, and a Master of Science in Computer Science from Queen’s University in Kingston, Ontario. During his graduate studies, he served as a Research Assistant in the RTL Lab at Queen’s University, where he contributed to research in networks and edge computing. His academic background encompasses software systems, algorithms, and 5G networks. Currently, he works as a Software Engineer at Amazon, where he focuses on designing and implementing scalable software solutions. His professional interests include cloud computing, systems architecture, and the development of high-performance, fault-tolerant applications.
\end{IEEEbiography}

\begin{IEEEbiography}[{\includegraphics[width=1in,height=1.25in,clip,keepaspectratio]{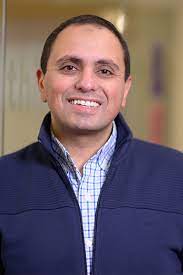}}]{Sameh Sorour (Late).} Dr. Sorour was an assistant professor at the School of Computing, Queen’s University, where he led several projects on autonomous and connected vehicles, edge intelligence, and wireless networks and services. He was a senior IEEE member and an Editor for IEEE Communications Letters. His research and educational interests lie in the broad areas of advanced computing, learning, and networking technologies for cyber-physical and autonomous systems. Dr. Sorour died in 2021. He was a wonderful scholar, instructor, mentor, and colleague; he will be greatly missed.

\end{IEEEbiography}

\begin{IEEEbiography}[{\includegraphics[width=1in,height=1.25in,clip,keepaspectratio]{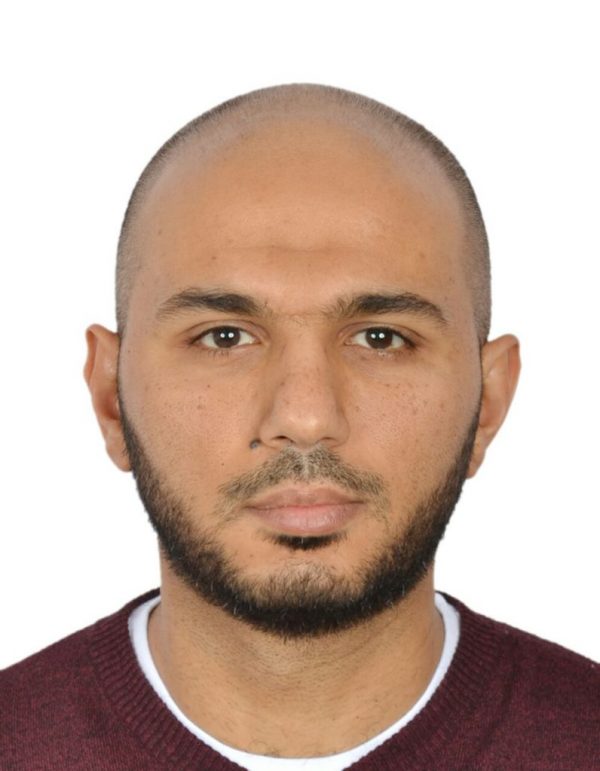}}]{Hesham ElSawy} (Senior Member, IEEE) an Associate Professor with the School of Computing, Queen's University, Kingston, ON, Canada. Prior to that, he was an assistant professor at King Fahd University of Petroleum and Minerals (KFUPM), Saudi Arabia, a Post-Doctoral Fellow at the King Abdullah University of Science and Technology (KAUST), Saudi Arabia, a Research Assistant at TRTech, Winnipeg, MB, Canada. He received the Ph.D. degree in electrical engineering from the University of Manitoba, Canada, in 2014.  He conducts research in the broad area of wireless communications and networking with a special focus on 5G/6G networks, Internet of Things, edge computing, non-terrestrial networks, and wireless security.  Dr. Elsawy is a recipient of the IEEE ComSoc Outstanding Young Researcher Award for Europe, Middle East, and Africa Region in 2018. He also received several best paper awards including the IEEE COMSOC Best Tutorial Paper Award in 2020 and IEEE COMSOC Best Survey Paper Award 2017. He is an Editor of the IEEE Transactions on Wireless Communications, the IEEE Transactions on Network Science and Engineering, and the IEEE Communications Letters. 
\end{IEEEbiography}

\begin{IEEEbiography}[{\includegraphics[width=1in,height=1.25in,clip,keepaspectratio]{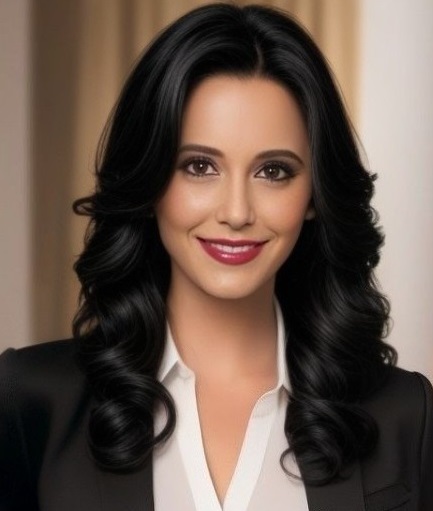}}]{Sara A. Elsayed }  received her PhD in Computer Science from the School of Computing at Queen’s University, Canada, in 2020. She is currently an Assistant Professor in the Department of Computer Science at the University of Calgary. Previously, Elsayed worked as a Postdoctoral Fellow and Adjunct Assistant Professor at Queen’s University, where she also managed and coordinated a large-scale research project focused on democratizing edge computing and edge intelligence in collaboration with Distributive, Ltd. In 2023, she received the Queen’s School of Computing (QSC) research award in recognition of her research contributions. Prior to joining Queen’s University, Elsayed was an Assistant Lecturer in the Faculty of Computers and Information, Cairo University, Egypt. She was the Co-founder and Program Manager of the Teaching \& Instructional Center (TIC), and was an Assistant Focal Point in the Support Office for Research Cooperation \& Mobility (SORCAM), Engineering Sector, Egypt. She is also a certified Cisco Networking Academy Instructor. Her research interests include Edge Computing, Edge Intelligence, Vehicular Networks, Caching, Internet of Things, Artificial Intelligence, and Intelligent Systems. She has several publications in top venues and is a member of the IEEE.
\end{IEEEbiography}

\begin{IEEEbiography}[{\includegraphics[width=1in,height=1.25in,clip,keepaspectratio]{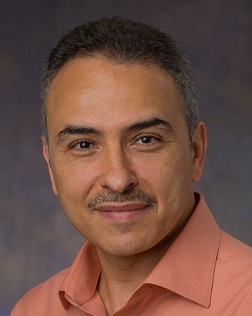}}]{Hossam S. Hassanein} (Fellow, IEEE) is currently
a leading Researcher in the areas of broadband,
wireless and mobile networks architecture, protocols, control, and performance evaluation. His record
spans more than 600 publications in journals, conferences, and book chapters, in addition to numerous
keynotes and plenary talks in flagship venues. He has
received several recognition and best paper awards at
top international conferences. He is the Founder and
the Director of the Telecommunications Research
Laboratory (TRL), School of Computing, Queen’s
University, with extensive international academic and industrial collaborations.
He is a recipient of the 2016 IEEE Communications Society Communications
Software Technical Achievement Award for outstanding contributions to
routing and deployment planning algorithms in wireless sensor networks and
the 2020 IEEE IoT, Ad Hoc and Sensor Networks Technical Achievement and
Recognition Award for significant contributions to technological advancement
of the Internet of Things, ad hoc networks, and sensing systems. He is the
former Chair of the IEEE Communication Society Technical Committee on
Ad hoc and Sensor Networks (TC AHSN). He is an IEEE Communications
Society Distinguished Speaker [a Distinguished Lecturer (2008–2010)].
\end{IEEEbiography}

\end{document}